\documentclass{ar-1col}
\pdfoutput=1

\usepackage{url}
\usepackage{graphicx}
\usepackage{float}
\usepackage{afterpage}
\usepackage{epsfig}
\usepackage{amssymb}
\usepackage{amsmath}
\usepackage{bm}
\usepackage{dsfont}
\usepackage{multirow}
\usepackage{xcolor}
\usepackage{ulem}
\usepackage{hyperref}
\usepackage{mcite}
\usepackage{cite}
\usepackage{booktabs}
\usepackage{atbegshi,picture}
\usepackage{lipsum}

\jname{Annu.Rev.Nucl.Part.Sci.}
\jvol{70}
\jyear{2020}
\doi{10.1146/annurev-nucl-011720-042725}

\AtBeginShipoutNext{\AtBeginShipoutUpperLeft{%
  \put(\dimexpr\paperwidth-0.5cm\relax,-1.5cm){\makebox[0pt][r]{Nikhef/2020-003}}%
}}

\begin{document}

% Page header
\markboth{Ethier and Nocera}{Parton Distributions in Nucleons and Nuclei}

% Title
\title{Parton Distributions\\ in Nucleons and Nuclei}

%Authors, affiliations address.
\author{Jacob J. Ethier,$^{1,2}$ and Emanuele R. Nocera$^{1,*}$
\affil{$^1$Nikhef Theory Group, Science Park 105, 1098 XG Amsterdam, 
The Netherlands}
\affil{$^2$Department of Physics and Astronomy, Vrije Universiteit Amsterdam,
1081 HV Amsterdam, The Netherlands}
\affil{$^*$email: e.nocera@nikhef.nl}}

%Abstract
\begin{abstract}
We review the current status of Parton Distribution 
Function (PDF) determinations for unpolarized and 
longitudinally polarized protons and for unpolarized 
nuclei, which are probed by high-energy hadronic 
scattering in perturbative Quantum Chromodynamics 
(QCD). We present the established theoretical 
framework, the experimental information, and the 
methodological aspects inherent to any modern PDF 
extraction. Furthermore, we summarize the present 
knowledge of PDFs and discuss their limitations 
in both accuracy and precision relevant to advance 
our understanding of QCD proton substructure and 
pursue our quest for precision in the Standard Model 
and beyond. In this respect, we highlight various
achievements, discuss contemporary issues in PDF 
analyses, and outline future directions of progress.
\end{abstract}

%Keywords

%Preprint number
\maketitle

%Table of Contents
\tableofcontents

%Body of the text
\vspace{-0.4cm}
\section{A BRIDGE BETWEEN LOW AND HIGH ENERGIES}
\label{sec:introduction}

Nucleons (protons and neutrons) are bound states that 
make up all nuclei, and hence most of the visible matter in the Universe. 
Unraveling their fundamental structure in terms of their elementary 
constituents --- quarks and gluons, or collectively called partons --- 
is currently one of the main challenges at the boundary of hadron and 
particle physics. Such an 
understanding is rooted in the theoretical framework provided by the 
Standard Model (SM), part of which contains the field theory 
that describes the strong interactions of color-charged partons:
Quantum Chromodynamics (QCD). 
Since energy grows with the separation between color charges, one
of the defining features of QCD, partons are confined to exist only in 
neutral color combinations called hadrons, among which are nucleons.

Nucleons are probed by scattering a beam of leptons or protons/antiprotons 
from them in large-momentum-transfer processes. Since the 
elementary interactions occur at distances much shorter than the confinement
scale (the scale at which partons are bound into nucleons), the measurable
cross section of the process can be determined by folding the partonic cross 
section with Parton Distribution Functions (PDFs). The former encodes the
scattering of quasi-free partons in terms of process-dependent kernels 
that are computed perturbatively in QCD, while the latter detail the 
momentum distribution of partons that enter the elementary 
scattering process in terms of universal functions.

Parton distributions are ubiquitous in hadron and particle physics: they are 
essential tools to interpret experimental data for a variety of 
hard-scattering processes in light of the underlying theory. Such processes are
measured with the greatest precision by different experiments at a number of 
facilities around the world, among which is the largest proton collider
ever built: the Large Hadron Collider (LHC). Collectively, they cover 
a wide range of momentum-transfer energies and are therefore sensitive 
to different aspects of the theory that are not yet fully understood. Among
these are the following outstanding questions: 
Are the Higgs boson dynamics the same as those prescribed by the SM, or are 
there other phenomena at work? What is the real origin of electroweak symmetry 
breaking and particles' mass? How do nucleon bound 
states emerge from parton interactions? 
How does the proton spin emerge from quarks and
gluons? How are parton dynamics modified within the bound nucleons of nuclei? 
What is the interplay with astrophysics?

To successfully address --- and possibly answer --- these questions, a careful
determination of PDFs and their uncertainties is mandatory. Currently, this
cannot be reliably achieved from first principles. Instead, PDFs are modeled
by means of some parameterization, which is then optimized by comparing the 
PDF-dependent prediction of one or more physical process to its actual 
measurement, a procedure that is called (global) QCD analysis (or fit). In 
this sense, PDF extractions can be labeled generally as a nonlinear regression 
problem, whereby one has to learn a set of functions from data. The PDFs can 
then be used to describe any other process that depends on them due to their
universal property.

The purpose of this review is twofold. First of all, it aims at providing an 
accessible, yet complete, introduction to the art of PDF fitting 
with a focus on the theoretical, experimental, and methodological
input that enters modern QCD analyses. In this respect, our review is more
concise and pedagogical than those recently presented 
in~\cite{Gao:2017yyd,Kovarik:2019xvh,Aidala:2012mv}.
Secondly, it aims at presenting
altogether various species of collinear PDFs: unpolarized PDFs, relevant for 
precision physics in the SM; helicity (or polarized) PDFs, relevant for the 
nucleon spin structure; and nuclear PDFs, relevant for the understanding of 
cold nuclear matter effects. In this respect, our review updates and extends 
those in~\cite{Forte:2013wc,Jimenez-Delgado:2013sma}, and presents a critical 
snapshot of PDFs as a bridge between low and high energy physics.

The review is organized as follows. In Sect.~\ref{sec:theory} we present the 
theoretical formalism used to study PDFs in perturbative QCD.
In Sect.~\ref{sec:expdata} we discuss the typical data sets included in a
global QCD analysis, and how they constrain the different parton flavors. 
We review the methodological aspects relevant
to PDF fitting and techniques used to represent 
their uncertainties in Sect.~\ref{sec:methodology}. 
Sect.~\ref{sec:results} is an overview
of state-of-the-art PDF fits, their main features, and phenomenological
implications. This is followed by a discussion on the relevance of PDFs to 
some of the 
questions outlined above, and by presenting a critical appraisal on future
prospects, in Sect.~\ref{sec:relevance}. Finally, we conclude with an outlook 
in Sect.~\ref{sec:outlook}.

\vspace{-0.4cm}
\section{THEORETICAL INPUT}
\label{sec:theory}

In this section we review the standard theoretical framework used to study PDFs 
in perturbative QCD. In particular, we describe the factorization of hadronic 
observables, the details of the perturbative computations, and the 
theoretical constraints that PDFs must fulfill.

\subsection{Factorization of hadronic observables}
\label{subsec:factorization}

Factorization theorems~\cite{Collins:1989gx} provide the framework in which 
perturbative QCD calculations can be performed for a class of sufficiently 
inclusive hadronic observables that are measured in large-momentum-transfer 
processes. In the region of asymptotic freedom~\mcite{Gross:1973id,
*Politzer:1973fx}, elementary QCD interactions occur at 
distances much shorter than the confinement scale --- roughly the order of the 
nucleon's quantum-mechanical wavelength --- and short-distance interactions of 
partons (expressed in terms of process-dependent 
partonic cross sections calculable in perturbative QCD) 
can be seperated from their long-distance momentum distributions 
(given by the non-perturbative, universal PDFs). Factorization realizes 
the convolution of these two parts to make predictions of 
experimental observables. While we refer to~\cite{Ellis:1991qj,
Collins:2011zzd} for a detailed treatment of the 
subject, in the following we briefly discuss how factorization behaves for 
lepto- and hadro-production, two classes of processes that are key to 
this review.

Deep-inelastic scattering (DIS), the high-energy inclusive scattering of 
leptons $\ell$ and nucleons $N$ ($\ell N\to\ell^\prime X$, where $X$ is 
unobserved), still plays a central role in discussing factorization, 
both because it was the first process to establish the 
existence of partons inside the nucleon, and because it comprises 
a majority of the experimental information used in PDF analyses (see 
Sect.~\ref{sec:expdata}). The unpolarized (polarized) DIS 
cross sections $\sigma$ ($\Delta\sigma$) for neutral current ($i$=NC)
interactions involving a photon or $Z^0$-boson exchange, and for
charged current ($i$=CC) interactions given by a $W^\pm$-boson
exchange, can be written as~\cite{Blumlein:2012bf}
%-------------------------------------------------------------------------------
\begin{eqnarray}
\frac{d^2\sigma^i}{dxdy}
&=&
\frac{2\pi\alpha^2}{xyQ^2}\eta^i
\left[
Y_+F_2^i\mp Y_-xF_3^i-y^2F_L^i
\right]
\,\mbox{,}
\label{eq:DISXsecunp}
\\
\frac{d^2\Delta\sigma^i}{dxdy}
&=&
\frac{4\pi\alpha^2}{xyQ^2}\eta^i
\left[
-Y_+g_4^i-\lambda_\ell Y_-2xg_1^i+y^2g_L^i
\right]
\,\mbox{,}
\label{eq:DISXsecpol}
\end{eqnarray}
%-------------------------------------------------------------------------------
where the standard DIS kinematic variables are: the momentum fraction 
$x=Q^2/2q\cdot P$, the inelasticity $y=q\cdot P/k\cdot P$, and the energy
$Q^2=-q^2$, with $k$, $P$, and $q$ being the lepton, nucleon, and transferred
momentum, respectively.
The polarized cross section is defined as the
difference $\Delta\sigma^i=\sigma^i(\lambda_N=-1,\lambda_\ell)
-\sigma^i(\lambda_N=+1,\lambda_\ell)$, where $\lambda_N$ ($\lambda_\ell$) 
is the nucleon (lepton) helicity which can have value 
$\pm1$ corresponding to its orientation parallel $(+)$ or antiparallel 
$(-)$ to the beam direction. 
The electroweak factors $\eta^i$ depend on the interaction type
and on the lepton helicity $\lambda_\ell$,
%-------------------------------------------------------------------------------
\begin{equation}
\eta^{\rm CC}
=(1\pm\lambda_\ell)^2\frac{1}{2}
\left(
\frac{G_FM_W^2}{4\pi\alpha}\frac{Q^2}{Q^2+M_W^2}
\right)^2
\mbox{\,,}
\qquad
\eta^{\rm NC}
= 1
\,\mbox{,}
\label{eq:ewfactors}
\end{equation}
%-------------------------------------------------------------------------------
where $G_F$ is the Fermi constant and $M_W$ the $W$-boson mass. 
In Eqs.~\eqref{eq:DISXsecunp} and \eqref{eq:DISXsecpol}, we have denoted the 
Quantum Electrodynamics (QED) coupling as $\alpha$, defined the kinematic 
factors $Y_\pm=1\pm(1-y)^2$, and neglected terms proportional to 
$M^2/Q^2$, with $M$ being the mass of the nucleon. In Eqs.~\eqref{eq:DISXsecunp}
and \eqref{eq:ewfactors} the $\mp$ ($\pm$) sign refers to either an incoming 
electron or neutrino ($-$), or to an incoming positron or antineutrino ($+$). 
At leading twist, the (un)polarized structure functions $F_2^i$, $F_3^i$, $F_L^i$
and $g_4^i$, $g_1^i$, $g_L^i$, collectively labeled $\mathcal{F}^i$, 
factorize as,
%-------------------------------------------------------------------------------
\begin{equation}
\mathcal{F}^i(x,Q^2)
=
x\sum_{\rm f}^{n_{\rm f}+1}\int_x^1\frac{dx^\prime}{x^\prime}
C_{\rm f}^i\left(\frac{x}{x^\prime},a_s(Q^2)\right){\mathrm f}^N(x,Q^2)
\,\mbox{,}
\label{eq:struct_fact}
\end{equation} 
%-------------------------------------------------------------------------------
where the sum runs over all active quarks $n_{\rm f}$ and the gluon, $C_{\rm f}^i$ 
are the partonic scattering coefficients computed as a power series in the 
strong coupling 
$a_s=\alpha_s/4\pi$, and ${\mathrm f^N}$ are either the unpolarized ($f^N$)
or the polarized ($\Delta f^N$) PDFs in a nucleon $N$. Finally, the PDFs 
can be further decomposed in terms of 
the net momentum density of partons
aligned ($\uparrow$) or anti-aligned ($\downarrow$) to the parent 
nucleon's polarization:
%-------------------------------------------------------------------------------
\begin{equation}
f^N=\frac{1}{2}\left(f^{N,\uparrow}+f^{N,\downarrow}\right)
\quad
\text{and}
\quad
\Delta f^N=\frac{1}{2}\left(f^{N,\uparrow}-f^{N,\downarrow}\right)
\,\mbox{.}
\label{eq:PDFs}
\end{equation}
%-------------------------------------------------------------------------------
The PDFs also have a bilocal operator
definition in canonical field theory (see Ref.~\cite{Collins:1989gx} and Sect.~2 in 
Ref.~\cite{Kovarik:2019xvh}). For example, the
unpolarized quark PDFs can be written as,
%-------------------------------------------------------------------------------
\begin{equation}
f^N=\frac{1}{4\pi}\int dy^- e^{-ixP^+ y^-}
\langle N | \bar\psi_f (0,y^-,{\bm 0})\gamma^+ W(y^-,0)\psi_f(0) | N \rangle
\,\mbox{,}
\label{eq:operatorial_def}
\end{equation}
%-------------------------------------------------------------------------------
where $\psi_f$, $\bar\psi_f$ are the quark fields, $\gamma$ are the Dirac
matrices, $W$ is the Wilson line ensuring the Gauge invariance of the 
operator, and all four-vectors have been expressed using light-cone 
coordinates. Similar expressions also hold for the gluon PDF and for polarized PDFs.

Additional particles in the final state 
can be measured in conjunction with the outgoing lepton 
in lepton-nucleon scattering. In 
semi-inclusive DIS (SIDIS), for instance, a single hadron type $h$ is 
identified 
($\ell N\to \ell^\prime h X$). In the case of unpolarized incoming particles, 
the SIDIS cross section reads~\cite{deFlorian:1997zj},
%-------------------------------------------------------------------------------
\begin{equation}
\frac{d^3\sigma^h}{dxdydz}
=
\frac{4\pi\alpha^2}{yQ^2}
\left[
Y_+ F_1^h + (1-y)F_L^h
\right]
\,\mbox{.}
\label{eq:SIDISXsec}
\end{equation}
The process is sufficiently inclusive to allow the structure functions 
$F_{1,L}^h$ to factorize as
\begin{equation}
F_{1,L}^h(x,z,Q^2)
=
\sum_{f,f^\prime}^{n_{f,f^\prime}+1}
\int_x^1\frac{dx^\prime}{x^\prime}\int_z^1\frac{dz^\prime}{z^\prime}
C_{ff^\prime}^h\left(\frac{x}{x^\prime}, \frac{z}{z^\prime},a_s(Q^2)\right)f^N(x,Q^2) 
D_{f^\prime}^h(z,Q^2)
\,\mbox{,}
\label{eq:SIDISfact}
\end{equation}
%-------------------------------------------------------------------------------
where the hard scattering coefficients are now denoted as $C_{f,f^\prime}^h$, 
and the fragmentation function (FF) $D_{f^\prime}^h$~\cite{Metz:2016swz} 
is introduced. This is the 
time-like counterpart of the PDF encoding the hadronization of a parton 
$f^\prime$ into a hadron $h$. The corresponding momentum fraction is defined as 
$z=P\cdot P^h/P\cdot q$, where $P^h$ is the four-momentum of the outgoing 
hadron.

The cross section for a generic unpolarized nucleon-nucleon process, 
which depends on a single scale $Q$ ({\it e.g.} $pp\to A X$, where $A$ can be,
for example, a single jet of hadrons, a lepton or quark pair, or an electroweak 
boson) also has a factorized expression,
%-------------------------------------------------------------------------------
\begin{equation}
\sigma_A(s,Q^2)
=
\sum_{f,f^\prime}^{n_{f,f^\prime}+1}\sigma^0_{ff^\prime}
\int_\tau^1\frac{dx^\prime}{x^\prime}
\int_{x^\prime}^1\frac{dx^{\prime\prime}}{x^{\prime\prime}}
f^N(x^{\prime\prime},Q^2)f^{\prime,N}\left(\frac{x^\prime}{x^{\prime\prime}},Q^2 \right) 
C_{ff^\prime}\left(\frac{\tau}{x^\prime},a_s(Q^2) \right)
\,\mbox{,}
\label{eq:pp_fact}
\end{equation}
%-------------------------------------------------------------------------------
where $s$ is the center-of-mass energy of the hadronic collision,  
$\tau=Q^2/s$ is the scaling variable of the hadronic process, 
$\sigma_{ff^\prime}^0$ is the leading order (LO) partonic cross section, $f^N$ and 
$f^{\prime,N}$ are the unpolarized PDFs stemming from each nucleon, and 
$C_{ff^\prime}$ are the hard scattering coefficients. 
Moreover, expressions similar to Eq.~\eqref{eq:pp_fact} exist for processes 
with one 
or both colliding nucleons polarized, for factorizable multi-scale
processes ({\it e.g.} Higgs production in $W$ fusion), for less inclusive 
processes ({\it e.g.} electroweak boson production in association with jets),
and for one-particle inclusive production ({\it e.g.} neutral pion production).
In the last case, an additional convolution with the final particle 
FF must be included. The factorized result of 
Eq.~\eqref{eq:pp_fact} can also be extended 
to total and differential cross sections.

Should hard scattering processes occur off nuclei instead of nucleons, it is 
custom to assume that they can still be described in 
terms of factorization theorems~\cite{Barshay:1975zz}. 
If so, the partonic scattering coefficients in
Eqs.~\eqref{eq:struct_fact}, \eqref{eq:SIDISfact}, and \eqref{eq:pp_fact} remain
the same but the PDFs are of nuclei, defined as the 
average of proton and neutron densities bound in a nucleus,
%-------------------------------------------------------------------------------
\begin{equation}
f^{(A,Z)}(x,Q^2)
=
\frac{Z}{A}f^{p/A}(x,Q^2)+\frac{A-Z}{A}f^{n/A}(x,Q^2)
\,\mbox{.}
\label{eq:nucPDFs}
\end{equation}
In this expression, the pair of atomic numbers $(A,Z)$ identifies the nuclear
isotope, and $f^{p,n/A}$ are the proton and neutron bound PDFs, usually related 
to the nucleon PDFs by,
%-------------------------------------------------------------------------------
\begin{equation}
f^{p,n/A}(x,Q^2)=R_f^A(x,Q^2) f^{p,n}(x,Q^2)
\,\mbox{.}
\label{eq:Rmodification}
\end{equation}
%-------------------------------------------------------------------------------
Since the nuclear and nucleon PDFs are defined by the same leading twist 
operators of Eq.~\eqref{eq:operatorial_def} (though acting on different states),
it is natural to assume that nuclear modifications can be absorbed into the 
PDFs without altering factorization theorems. This assumption might not hold 
for processes or kinematic regions subject to sizable higher-twist 
corrections~\cite{Kovarik:2019xvh}.

\subsection{Perturbative calculations}
\label{subsec:perturbative}

The partonic scattering coefficients $C_{\rm f}^i$, $C_{ff^\prime}^h$ and
 $C_{ff^\prime}$ entering Eqs.~\eqref{eq:struct_fact}, \eqref{eq:SIDISfact} 
and~\eqref{eq:pp_fact}, which we collectively denote as $\mathcal{C}$, 
can be expressed as a perturbative series in the strong coupling,
%-------------------------------------------------------------------------------
\begin{equation}
\mathcal{C}=\sum_{k=0}a_s^kc^{(k)}
\,\mbox{,}
\label{eq:coeff_pert}
\end{equation}
%-------------------------------------------------------------------------------
where the explicit form of the coefficients $c^{(k)}$ depend on the specific 
process. At the lowest perturbative order, LO, they either vanish or 
are proportional to a Dirac delta, in which case the hadronic cross sections 
reduce to a combination of the PDFs that couple to the relevant final state. 
Each process is therefore sensitive to different partons 
(see Sect.~\ref{sec:expdata}).

The computation of partonic cross sections to higher orders, usually 
next-to leading and next-to-next-to-leading orders (NLO and NNLO), generally 
entail three classes of singularities: ultraviolet (UV) singularities, 
which are renormalized through the 
running of the QCD coupling $a_s(\mu_R)$, at a scale $\mu_R$; infrared 
singularities associated with loop graphs, which cancel 
corresponding soft singularities from the emission of real gluons; and collinear 
singularities due to the initial partons emitting gluons at zero angle,
which are subtracted by terms arising in the renormalization of the 
PDF operators~\cite{Collins:1989gx}. The resulting partonic cross sections
are therefore free of any infrared type singularities, but remain dependent on
a renormalization scale $\mu_R$ as a result of regularizing the UV divergences.
Renormalization also implies the choice of a scheme, the one
most widely used being the modified minimal subtraction scheme 
($\overline{\rm MS}$)~\cite{tHooft:1972tcz,*tHooft:1973mfk,*Bardeen:1978yd},
which includes additional common factors in the counterterms required 
to make finite predictions. 

Since the PDFs defined in factorization 
are renormalized separately, an additional scale dependence 
$\mu_F$ arises in their argument, 
the choice of which can be made independently 
from the scale choice in the partonic cross section calculation.
However, both the renormalization and the factorization scales $\mu_R$ and $\mu_F$ 
are unphysical, {\it i.e.} the hadronic cross section does not depend on them 
should it be computed to all orders in perturbation theory. At a given
fixed order $a_s^k$ these cancellations will only be approximate, that is, 
%-------------------------------------------------------------------------------
\begin{equation}
\mu_F^2\frac{\partial\sigma}{\partial\mu_F^2}=0+\mathcal{O}(a_s^{k+1})
\qquad
\text{and}
\qquad
\mu_R^2\frac{\partial\sigma}{\partial\mu_R^2}=0+\mathcal{O}(a_s^{k+1})
\label{eq:MIssHO}
\end{equation}
for the hadronic cross section $\sigma$ of any of the processes in 
Eqs.~\eqref{eq:DISXsecunp},\eqref{eq:DISXsecpol},\eqref{eq:SIDISXsec} 
and~\eqref{eq:pp_fact}. This suggests that the dependence of the 
hadronic cross section on the renormalization and factorization scale choice 
decreases as one carries out the perturbative calculation to higher orders 
(unless new kinematic configurations open up), and that scale variations can 
be used to estimate the accuracy of the perturbative order. 
Unless otherwise stated, we assume $\mu_R=\mu_F=Q$ in this review 
(including the previous section).

The dependence of the strong coupling on the renormalization scale and of the
PDFs on the factorization scale is completely determined by the invariance of 
the renormalization group and the corresponding equations: the QCD beta 
function for the strong coupling (which is currently known up to five 
loops~\cite{Herzog:2017ohr,*Luthe:2017ttg,*Herzog:2018kwj}), and the DGLAP 
evolution equations~\cite{Gribov:1972ri,*Lipatov:1974qm,*Altarelli:1977zs,
*Dokshitzer:1977sg} for the PDFs. This is a set of $2n_{\rm f}+1$ coupled
integro-differential equations of the form,
%-------------------------------------------------------------------------------
\begin{equation}
Q^2\frac{\partial{\rm f}(x,Q^2)}{\partial Q^2}
=
\sum_{\rm f}^{2n_{\rm f}+1}\int_x^1\frac{dx^\prime}{x^\prime}
\mathcal{P}_{{\rm f}^\prime {\rm f}}\left(\frac{x}{x^\prime},Q^2 \right)
{\rm f}^N(x^\prime,Q^2)
\,\mbox{,}
\label{eq:DGLAP}
\end{equation}
%-------------------------------------------------------------------------------
often conveniently rewritten in terms of the gluon, non-singlet 
${\rm f}_{\rm NS}^N={\rm q}-\bar{\rm q}$ and singlet 
${\rm f}_{\rm S}^N=\sum_q^{n_{\rm f}}{\rm q}^+$ (with 
${\rm q}^+={\rm q}+\bar{{\rm q}}$) quark-antiquark combinations. 
In this way, Eq.~\eqref{eq:DGLAP} reduces to 
$2n_{\rm f}-1$ decoupled equations for the non-singlet distributions and two
coupled equations for the gluon and singlet distributions. 
The splitting functions $\mathcal{P}_{{\rm f}^\prime {\rm f}}$
can be either unpolarized ($P_{{\rm f}^\prime {\rm f}}$) or polarized 
($\Delta P_{{\rm f}^\prime {\rm f}}$) and can be expanded as a power series in $a_s$,
%-------------------------------------------------------------------------------
\begin{equation}
\mathcal{P}_{{\rm f}^\prime {\rm f}}=\sum_{k=0}a_s^{k+1}p^{(k)}
\,\mbox{,}
\label{eq:splitting}
\end{equation}
%-------------------------------------------------------------------------------
with the coefficients $p^{(k)}$ computed in perturbative QCD up to 
NNLO~\cite{Moch:2004pa,*Vogt:2004mw} and partly to 
N$^3$LO~\cite{Ablinger:2017tan,*Moch:2017uml} for $P_{{\rm f}^\prime{\rm f}}$, 
and NNLO for $\Delta P_{{\rm f}^\prime{\rm f}}$~\cite{Moch:2015usa,*Vogt:2008yw}.
Expressions similar to Eqs.~\eqref{eq:DGLAP}-\eqref{eq:splitting} also hold
for the FF in Eq.~\eqref{eq:SIDISfact}, with the corresponding splitting
functions known up to NNLO~\cite{Mitov:2006ic,*Moch:2007tx,*Almasy:2011eq}.
Since the coefficients for partonic cross sections in 
Eq.~\eqref{eq:coeff_pert} are known up to NNLO for most of the unpolarized
processes and up to NLO for most of the polarized processes, the corresponding 
PDFs can be determined only up to these accuracies. 

Beyond LO, the coefficients of partonic 
scattering in Eq.~\eqref{eq:coeff_pert} and
of the splitting functions in Eq.~\eqref{eq:splitting}
contain terms proportional to $\ln Q^2$, $\ln (1/x)$ and $\ln (1-x)$. 
While the first
appears in the splitting kernels and are summed by the DGLAP equations, 
resummation techniques~\cite{Ciafaloni:2003ek,*Ciafaloni:2003rd} 
and BFKL equations~\cite{Fadin:1975cb,*Kuraev:1976ge,*Kuraev:1977fs,
*Balitsky:1978ic} may be used to deal with $\ln (1/x)$ terms 
at small $x$. At large $x$, threshold resummation 
techniques~\cite{Catani:2014uta} handle terms proportional to $\ln (1-x)$.

Sums over the number of flavors $n_{\rm f}$ appear in all factorization 
formulas and in the evolution equations. Decoupling 
arguments~\cite{Appelquist:1974tg} imply that the contribution of heavy quark 
flavors to any process are power-suppressed at scales which are below the 
threshold for their production~\cite{Collins:1978wz}. Therefore, when 
expressing predictions for processes at
different scales in terms of the same PDF set, it is necessary to use a
variable-flavor number (VFN) scheme in which different numbers of
active flavors are adopted consistently. In the vicinity of the threshold for
heavy quark production, the quark mass $m$ cannot be ignored. This can be 
accounted for in a general-mass VFN (GM-VFN) scheme that interpolates, in a 
model-dependent way, between the fixed-flavor number (FFN) scheme near 
production threshold and the asymptotic result of the zero-mass VFN (ZM-VFN).
In the FFN scheme, heavy quark mass effects are built into the 
partonic cross sections, but terms proportional to large logarithms of the form 
$\ln(Q/m)$ are not resummed; in the ZM-VFN scheme, heavy quark mass effects are 
ignored, but $\ln(Q/m)$ terms are instead resummed into the heavy quark PDFs. 
Massive coefficient functions are known up to $\mathcal{O}(\alpha_s^2)$ for 
both neutral-~\cite{Laenen:1992zk,*Laenen:1992xs} and 
charged-current~\cite{Berger:2016inr,*Gao:2017kkx} DIS.
A number of partial results also exist at $\mathcal{O}(\alpha_s^3)$,
see~\cite{Ablinger:2017err,*Ablinger:2019gpu} and references therein. Various
GM-VFN schemes have been worked out in the literature up to NNLO,
including the exact dependence on the heavy-quark mass up to 
$\mathcal{O}(\alpha_s^2)$: ACOT~\cite{Aivazis:1993pi,*Collins:1998rz,
*Guzzi:2011ew}, TR~\cite{Thorne:1997ga,*Thorne:1997uu,*Thorne:2006qt} 
and FONLL~\cite{Cacciari:1998it,*Forte:2010ta}.
These schemes differ by subleading terms, which may not be entirely negligible 
at NLO in the vicinity of the quark threshold, but rapidly decrease at 
NNLO~\cite{Binoth:2010nha}. In the case of FONLL, the GM-VFNS was 
generalized to include mass effects for heavy-quark initiated 
contributions~\cite{Ball:2015tna,*Ball:2015dpa,*Forte:2019hjc} should 
non-perturbative charm and bottom quark content in the proton be required. 
While most of the details have been worked out in the unpolarized case, these 
can be extended to polarized PDFs and even FFs~\cite{Epele:2018ewr}.

We shall finally note that any process involving electroweak interactions
also receives higher-order electromagnetic or weak corrections. Both 
coefficent functions in Eq.~\eqref{eq:coeff_pert} and splitting 
functions in Eq.~\eqref{eq:splitting} should be modified, and additional PDFs 
for the photon (and generally for leptons) should be included in factorization
formulas and evolution equations~\cite{Bertone:2015lqa}. 
Since the QED coupling at the electroweak scale is 
$\alpha\sim\alpha_s^2\sim 0.01$, one expects NLO corrections
in the electromagnetic interaction to be roughly of the same order of NNLO QCD
corrections. Electroweak corrections have never been systematically included in 
global analyses of PDFs so far, except for unpolarized photon-initiated 
partonic cross sections. In this case, it was 
demonstrated~\cite{Manohar:2016nzj,*Manohar:2017eqh} that the photon PDF can be 
almost completely determined (to a precision comparable to quark PDFs) by 
relating the entire photon contribution to the proton structure function 
(refer to Sect.~7 in~\cite{Gao:2017yyd} for a thorough review).

\subsection{Theoretical constraints}
\label{subsec:evolution}

There are several theoretical constraints that can be used to help determine 
PDFs.
\vspace{-0.3cm}
\paragraph*{Positivity of hadronic observables.}
Hadronic observables must always be positive regardless of the shape or
size of the PDFs, which can be negative for unpolarized PDFs beyond the
lowest perturbative order, and for polarized PDFs at all orders. Positivity can 
be enforced by choosing {\it ad hoc} PDF parameterizations or imposed by 
defining a set of control cross sections on a grid of 
kinematics that captures a sufficiently large region of the phase space.
The parameter configurations that lead to negative cross sections are
subsequently discarded.
In principle, one could also isolate analytically the terms that lead to 
negative results from collinear subtraction in the DGLAP equations.
Such a procedure was used to derive positivity bounds on polarized PDFs from
their unpolarized counterparts~\cite{Altarelli:1998gn}.

\vspace{-0.3cm}
\paragraph*{Sum rules.}
Energy conservation implies that unpolarized parton momentum fractions $x$ 
carried by each parton must sum to unity, {\it i.e.} PDFs must fulfill the 
momentum sum rule,
%-------------------------------------------------------------------------------
\begin{equation}
\int_0^1 dx \sum_{f=q,\bar{q},g} xf(x) = 1
\,\mbox{.}
\label{eq:MSR}
\end{equation}
%-------------------------------------------------------------------------------
In addition, valence sum rules ensure that an unpolarized hadron maintains its valence
structure, {\it i.e.} for a proton made of two up quarks and one down quark,
%-------------------------------------------------------------------------------
\begin{equation}
\int_0^1 dx \left (u - \bar{u}\right) = 2
\,\mbox{,}\quad
\int_0^1 dx \left (d - \bar{d}\right) = 1
\,\mbox{,}\quad
\int_0^1 dx \left (s -\bar{s}\right) = 0
\,\mbox{.} 
\label{eq:VSR}
\end{equation}
%-------------------------------------------------------------------------------

\vspace{-0.3cm}
\paragraph*{{\rm\bf SU$_f$} symmetry relations.}
There are constraints that can be imposed on polarized PDFs using an 
established relationship between weak baryon decays and the non-singlet 
combinations of PDF moments. The first follows from the Bjorken sum 
rule~\mcite{Bjorken:1966jh,*Bjorken:1969mm}, which relates the 
difference between the lowest moments of the proton and neutron structure functions 
$g_1^p$ and $g_1^n$ to the isovector axial charge $g_A$ assuming exact
SU$_f$(2) symmetry. The second follows from relating the non-singlet 
combination of PDFs to the octet axial charge $a_8$, taken from weak hyperon 
decay assuming exact SU$_f$(3) symmetry. In terms of PDF moments,
%-------------------------------------------------------------------------------
\begin{equation}
\int_0^1 dx \left(\Delta u^+ - \Delta d^+\right) = g_A
\,\mbox{,}
\quad
\int_0^1 dx \left(\Delta u^+ + \Delta d^+ - 2\Delta s^+\right) = a_8
\,\mbox{.}
\label{eq:gA}
\end{equation}
%-------------------------------------------------------------------------------
The second relationship in particular
plays a vital role in constraining the strange polarization, forcing
its first moment to have a value $\sim-0.1$. While 
SU$_f$(2) symmetry has been confirmed in a recent QCD analysis
to almost 2\%~\cite{Ethier:2017zbq}, there is some uncertainty surrounding the 
level of SU$_f$(3) symmetry breaking in the value of 
$a_8$~\mcite{Jaffe:1989jz,*Bass:2009ed}. As a result, imposing the hyperon 
decay value in a fit can bias the shape of the strange polarization 
(see Sect.~\ref{sec:results}). 

\vspace{-0.3cm}
\paragraph*{Nuclear boundary condition.}
In addition to momentum and valence sum rules, nuclear PDFs for low $A$ nuclei
are constrained by unpolarized proton PDFs at $A=1$. This boundary condition 
is usually implemented by defining a nuclear PDF parameterization that
exactly reproduces the central value of the proton PDF at $A=1$.
Alternatively~\cite{AbdulKhalek:2019mzd}, one can fit the $A=1$ distribution 
along with other nuclei, and then discard parameter configurations that deviate 
from a given proton boundary condition (with its uncertainties and 
correlations). This approach results in smaller uncertainties for neighboring 
nuclei.  

\vspace{-0.4cm}
\section{EXPERIMENTAL INPUT}
\label{sec:expdata}

Theoretical predictions for hadronic cross sections, obtained by 
the formalism presented in Sect.~\ref{sec:theory}, must be compared to 
experimental data in order to determine PDFs. Table~\ref{tab:kinematics}
presents a summary of the main hadronic processes used in 
current global QCD analyses. For each reaction, we indicate the 
leading partonic process contributing 
to the corresponding factorization formulas, the probed parton flavors, and whether
there exists available data in the unpolarized, polarized, and nuclear sectors.
Furthermore, hadronic processes are grouped into seven 
categories: fixed target DIS, collider DIS, fixed target SIDIS, fixed target
Drell Yan (DY), collider DY, jet and hadron production, and top production. 

\begin{figure}[!t]
\centering
\includegraphics[width=\textwidth,clip=true,trim=0 0.8cm 0 0]{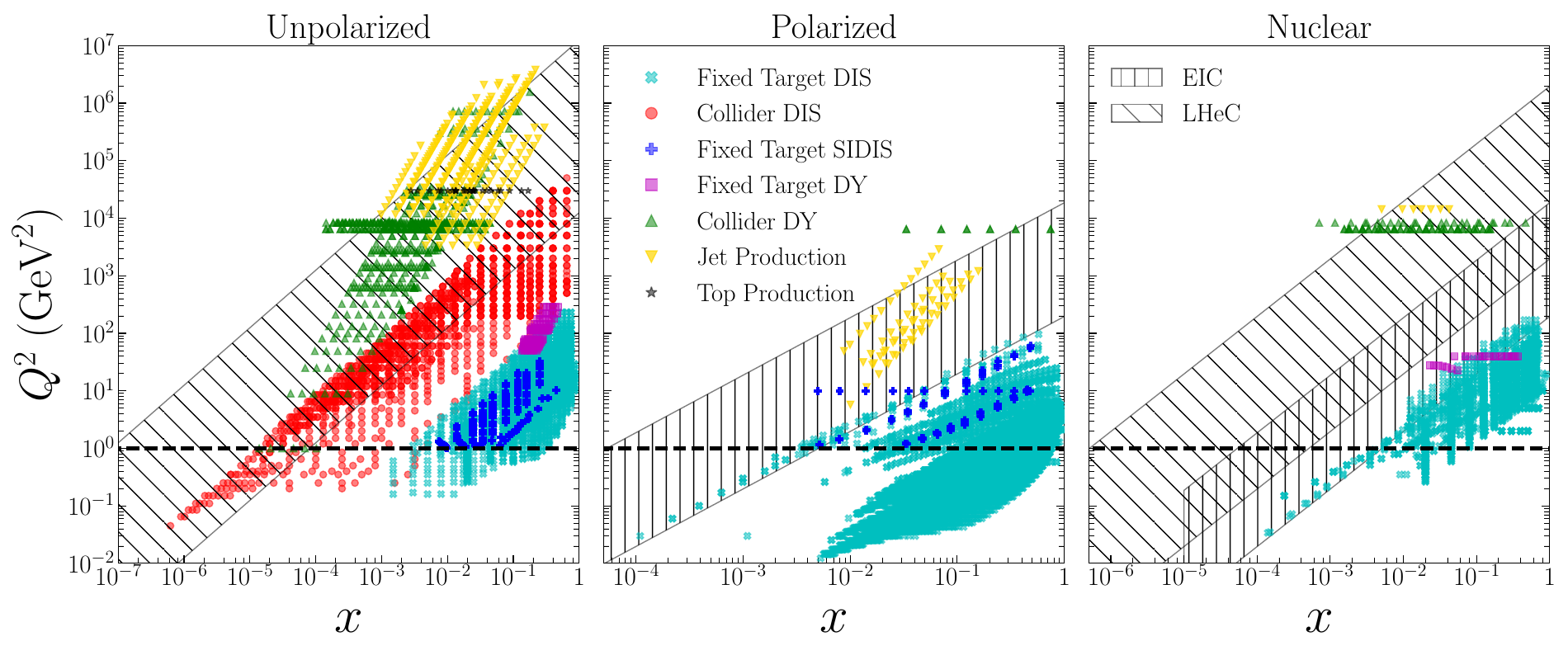}
\caption{The kinematic coverage, in the $(x,Q^2)$ plane, of the hadronic cross 
section data for the processes commonly included in global QCD analyses of 
unpolarized, polarized, and nuclear PDFs. The extended kinematic ranges attained 
by the LHeC and the EIC are also displayed.}
\label{fig:kinematics}
\end{figure}

The kinematic coverage in $x$ and $Q^2$ of the corresponding measurements
is displayed in Fig.~\ref{fig:kinematics} for each PDF species. 
Applicability of perturbative QCD implies that only the data above $Q^2=1$~GeV$^2$
are usually considered (indicated by the dashed horizontal line in Fig.~\ref{fig:kinematics}). 
Since each process probes a different parton flavor (or combinations thereof) in 
different kinematic regions, a global QCD analysis is required to piece the
information together. The ability to describe such a variety of hadronic 
observables simultaneously is in fact a great success of perturbative QCD.
Restricted datasets are also sometimes considered
in order to maximize consistency across measurements. However, this 
may limit the precision and accuracy of the resulting PDF distributions. 

%-------------------------------------------------------------------------------
\begin{table}[!t]
\centering
\begin{tabular}{lllccc}
\toprule
Hadronic Process & Partonic Process & Probed Partons & U & P & N\\  
\midrule
Fixed Target DIS & & & & & \\
$\ell^\pm\{p,n\}\to\ell^\pm + X$ 
   & $\gamma^*q \to q$ 
   & $\rm q^+,q,\bar{q},g$ 
   & \checkmark & \checkmark & \checkmark
\\
$\ell^\pm\{n,A\}/p \to \ell^\pm + X$
   & $\gamma^* d/u \to d/u$ 
   & $\rm d/u$  
   & \checkmark &            & \checkmark
\\
$\nu(\bar{\nu})A \to \mu^-(\mu^+) + X$
   & $W^*q\to q^\prime$
   & $\rm q,\bar{q}$
   & \checkmark &            & \checkmark
\\
$\nu A \to  \mu^-\mu^+ + X$
   & $W^* s \to c$
   & $\rm s$
   & \checkmark &            & \checkmark
\\
$\bar{\nu} A \to  \mu^+\mu^- + X$
   & $W^*\bar{s} \to \bar{c}$
   & $\rm\bar{s}$
   & \checkmark &            & \checkmark
\\
\midrule
Collider DIS & & & & &\\
$e^\pm p \to e^\pm + X$
   & $\gamma^* q \to q$
   & $\rm g,q,\bar{q}$
   & \checkmark & & 
\\
$e^+ p \to \bar{\nu} + X$
   & $W^+\{d,s\}\to \{u,c\}$
   & $\rm d,s$
   & \checkmark & & 
\\
$e^\pm p \to e^\pm c\bar{c} + X$
   & $\gamma^* c\to c, \gamma^*g\to c\bar{c}$
   & $\rm c,g$
   & \checkmark & & 
\\
$e^\pm p \to {\rm (di-)jet(s)} + X$
   & $\gamma^*g \to q \bar{q}$
   & $\rm g$
   & \checkmark & & 
\\
\midrule
Fixed Target SIDIS & & & & &\\
$\ell^\pm\{p,d\}\to\ell^\pm+h+X$
   & $\gamma^* q\to q$
   & $\rm u,\bar{u},d,\bar{d},g$
  & \checkmark & \checkmark
\\
$\ell^\pm\{p,d\}\to\ell^\pm c\bar{c}\to\ell^\pm D+X$
   & $\gamma^* g \to c\bar{c}$
   & $\rm g$
   &           & \checkmark &
\\
\midrule
Fixed Target DY & & & & &\\
$pp\to \mu^+\mu^- + X$
   & $u\bar{u}, d\bar{d} \to \gamma^*$
   & $\rm \bar{q}$
   & \checkmark & & 
\\
$p\{n,A\}/pp\to \mu^+\mu^- + X$
   & $(u\bar{d})/(u\bar{u}) \to \gamma^*$
   & $\rm \bar{d}/\bar{u}$
   & \checkmark &            & \checkmark
\\
\midrule
Collider DY & & & & &\\
$p\bar{p} \to (W^\pm\to\ell^\pm\nu) + X$
   & $ud\to W^+$, $\bar{u}\bar{d}\to W^-$
   & $\rm u,d,\bar{u},\bar{d}$
   & \checkmark & & 
\\
$p\{p,A\} \to (W^\pm\to\ell^\pm\nu) + X$
   & $u\bar{d}\to W^+$, $d\bar{u}\to W^-$
   & $\rm u,d,\bar{u},\bar{d}$
   & \checkmark & \checkmark & \checkmark
\\     
$p\bar{p}(p\{p,A\}) \to (Z\to\ell^+\ell^-) + X$
   & $uu,dd (u\bar{u}, d\bar{d}) \to Z$
   & $\rm u,d,g$
   & \checkmark & \checkmark & \checkmark
\\
$pp\to (W+c) + X$
   & $gs \to W^-c$, $g\bar{s}\to W^+\bar{c}$
   & $\rm s,\bar{s},g$
   & \checkmark & &
\\
$pp\to (\gamma^*\to\ell^+\ell^-)X$
   & $u\bar{u},d\bar{d}\to \gamma^*$,$u\gamma,d\gamma\to \gamma^*$
   & $\rm \bar{q},g,\gamma$
   & \checkmark & &
\\
\midrule
Jet and hadron production & & & & &\\
$p\bar{p}(p\{p,A\}) \to {\rm (di-)jet(s)} + X$
   & $gg,qg,qq\to {\rm jet(s)}$
   & $g,q$
   & \checkmark & \checkmark & \checkmark
\\
$p\bar{p}(pp) \to h + X$
   & $gg,qg,qq\to \pi,K,D$
   & $g,q$
   & \checkmark & \checkmark & 
\\
\midrule
Top Production & & & & &\\
$pp\to t\bar{t} + X$
   & $gg \to t\bar{t}$
   & $g$
   & \checkmark & & 
\\
$pp\to t+X$
   & $W^*q\to q^\prime$
   & $\rm q,\bar{q}$
   & \checkmark & & 
\\
\bottomrule
\end{tabular}
\\
\vspace{0.3cm}
\caption{The main hadronic processes commonly used to determine PDFs. For each
reaction, the leading partonic process, the probed partons,
and whether available data exists in the unpolarized (U), polarized (P), and 
nuclear (N) cases is indicated.}
\label{tab:kinematics}
\end{table}
%-------------------------------------------------------------------------------

%-------------------------------------------------------------------------------
%-------------------------------------------------------------------------------

As is apparent in Fig.~\ref{fig:kinematics}, the amount of experimental data
points and the variety of hadronic measurements is significantly 
larger for unpolarized proton scattering than for polarized and 
nuclear reactions. The reason for this
is that, historically, accelerating polarized nucleons and heavy 
ions has been technically more challenging. As a result, the experimental 
facilities pivotal for the extraction of unpolarized proton PDFs (namely the 
HERA collider and the LHC) have largely remained unparalleled in the 
polarized and nuclear cases. However, several new facilities and
upgrades are expected in the near future that aim to
decrease this discrepancy (see Sect.~\ref{sec:relevance}).

Nevertheless, in all cases, the bulk of the experimental information is 
provided by DIS. In the unpolarized sector, neutral and
charged current DIS has been measured by various experiments at the CERN
Super-Synchrotron (CERN-SPS), such as NMC, BCDMS and CHORUS, at the 
Stanford Linear Accelerator (SLAC), at Fermilab (NuTeV) and at HERA.
In the polarized case, neutral current DIS has been measured by 
various experiments at the CERN-SPS (EMC, SMC, COMPASS), at SLAC,
and at JLab. Lastly, in the nuclear case, 
both neutral and charged current DIS has been 
measured by experiments at the CERN-SPS, SLAC and Fermilab.
In principle, DIS is able to constrain all partons:
quark-antiquark total and valence combinations for each flavor through 
neutral and charged current interactions, respectively, and the gluon
through scaling violations via DGLAP evolution and higher order 
QCD corrections in the structure functions.

Complimentary to DIS are DY and $W^\pm$/$Z^0$ production measurements 
which are necessary to disentangle the sea quark distributions.
Various measurements of both fixed-target and collider DY scattering
have been performed: in the unpolarized case 
by the E665 and E866 experiments at Fermilab, by the D0 
and CDF experiments at the Tevatron, and by the ATLAS, CMS and LHCb experiments 
at the LHC; in the polarized case, by the STAR and 
PHENIX experiments at RHIC; and in the nuclear case, by the E772, E866, E605 
experiments at Fermilab. Sea quark distributions can likewise be probed in 
SIDIS, although analyses of the corresponding data are complicated by the 
non-perturbative final-state FF that enters the theoretical prediction 
(Eq.~\eqref{eq:SIDISfact}). Semi-inclusive pion, kaon, and charged hadron 
production were measured in both the unpolarized and polarized cases 
by the COMPASS experiment at CERN and by the HERMES experiment at HERA.

For the gluon distribution, jet and hadron production hold much of the constraining power.
In the unpolarized sector, jet production has been measured both in DIS by the H1
and ZEUS experiments at HERA, in proton-(anti)proton collisions by the D0 
and CDF experiments at the Tevatron, and by the ATLAS, CMS and ALICE experiments
at the LHC. In the polarized case, jet and pion production has been measured in 
proton-proton collisions by the STAR and PHENIX experiments at RHIC. In the
nuclear case, jet production and hadron production have been measured 
in proton-ion collisions by the CMS and ALICE experiments at the LHC
and by the STAR and PHENIX experiments at RHIC. Finally, single top 
and top pair production is used to probe the sea quark and 
gluon distributions, respectively. In this case, measurements are only available 
for unpolarized PDF analyses from the ATLAS and CMS experiments 
at the LHC. 

The interested reader can obtain further details on 
all of these measurements (including their references) from Sect.~3
of Ref.~\cite{Gao:2017yyd} for unpolarized PDFs, from Sect.~3 of
Ref.~\cite{Aidala:2012mv} for the polarized distributions, and from 
Sect.~3 of Ref.~\cite{Eskola:2016oht} for the nuclear PDFs.
The impact of these observables on PDFs will be discussed 
in Sect.~\ref{sec:results}.

\vspace{-0.4cm}
\section{METHODOLOGICAL INPUT}
\label{sec:methodology}

Since parton distributions cannot be computed reliably from first principles,
they are instead determined by comparing theoretical predictions
of hadronic cross sections in the form of 
Eqs.~\eqref{eq:struct_fact},\eqref{eq:SIDISfact} and~\eqref{eq:pp_fact} to
experimental data. In other words, PDFs must be modeled by a function of fit 
parameters and optimized by data to yield matching theoretical predictions. 
It can be classified more generally as a nonlinear regression problem, in which
a confidence interval in the space of PDFs is determined by optimizing
a suitable goodness-of-fit measure~\cite{Giele:2001mr}.
In this section, we discuss the methodological aspects to perform a global
QCD anaylsis, in particular the PDF parameterization, optimization, 
and uncertainty representation. 

\subsection{Parameterization}
\label{subsec:parameterization}

Unpolarized and polarized proton PDFs are usually parameterized for each 
parton ${\rm f}$ as,
%-------------------------------------------------------------------------------
\begin{equation}
{\rm f}^p(x,Q_0^2) 
= 
{\cal N}x^{\alpha_{\rm f}} (1-x)^{\beta_{\rm f}} {\cal I}(x ; {\bm a})
\,\mbox{,}
\label{eq:PDFparam}
\end{equation}
%-------------------------------------------------------------------------------
at an input scale $Q_0^2$. The power-like factors $x^{\alpha_{\rm f}}$ and 
$(1-x)^{\beta_{\rm f}}$ describe the low-$x$ and high-$x$ behavior of PDFs, 
and are inspired by general QCD arguments, namely Regge 
theory~\cite{Regge:1959mz} and Brodsky-Farrar quark counting 
rules~\cite{Brodsky:1973kr} for the small- and large-$x$
asymptotic limits, respectively. While various models can predict approximate values for 
$\alpha_{\rm f}$ and $\beta_{\rm f}$~\cite{Ball:2016spl,*Nocera:2014uea}, 
they are determined from the data in global QCD analyses. 
The factor $\mathcal{N}$ is an overall normalization
that accounts for theoretical constraints given by Eqs.~\eqref{eq:MSR}-\eqref{eq:gA},
which also imply $\beta_{\rm f}>0$ (to ensure PDF integrability).
The function ${\cal I}(x ; {\bm a})$, which depends on a set of parameters 
${\bm a}$, is defined to interpolate between the small- and large-$x$ regions. 
Typically it is chosen to have a simple polynomial form, {\it e.g.} 
${\cal I}(x) = (1+\gamma \sqrt{x}+\delta x+\ldots)$, where 
${\bm a}=\{\gamma,\delta,\dots\}$ are parameters to be determined by the fit, 
although some have used more flexible
Chebyshev~\cite{Harland-Lang:2014zoa,Dulat:2015mca} or Bernstein 
polynomials~\cite{Glazov:2010bw}. The function $\mathcal{I}$ can also be 
defined as the output of a neural network~\cite{Ball:2017nwa,Ball:2013lla}, 
to reduce bias coming from the choice of parameterization as much as possible.
In this context, neural networks provide a convenient unbiased set of 
(nonlinear) basis functions.

For PDFs of nuclei, an additional dependence on the atomic mass $A$ is required.
The treatment of such dependence is subject to various choices.
For instance, the nuclear modification $R_f^A(x,Q_0)$ 
can be parameterized directly~\cite{Eskola:2016oht}, in which case the
PDFs of a proton 
bound in a nucleus can  be constructed
according to Eq.~\eqref{eq:Rmodification}, given a proton PDF set. 
Alternatively, the nuclear proton PDF 
can be parameterized similarly to the free proton
PDFs~\cite{Kovarik:2015cma} by giving an $A$-dependent functional 
form to the parameters in Eq.~\ref{eq:PDFparam}, 
$(\alpha_f,\beta_f,{\bm a}) \to (\alpha_f(A),\beta_f(A),{\bm a}(A))$. 
Lastly, one can remain agnostic about the $A$ 
dependence by modeling it as part of a neural network~\cite{AbdulKhalek:2019mzd}. 

The choice of the input scale $Q_0$ is typically set below the charm mass 
threshold $m_c$ if heavier quark PDFs are assumed to be generated by QCD 
radiation, or slightly above it if the charm PDF is parameterized on the same 
footing as the light quark PDFs. Should the FFN scheme be 
chosen~\cite{Alekhin:2017kpj}, various input scales are used above each threshold 
leading to different PDF sets for every number of active flavors.
For unpolarized PDFs, there are usually seven
independent PDFs to parameterize, $f={u,\bar{u},d,\bar{d},s,\bar{s},g}$, 
which increase to eight or nine if the charm~\cite{Ball:2016neh} 
and the photon~\cite{Bertone:2017bme} PDFs are also parameterized. Should a 
restricted data set be used, or different species of PDFs be fitted, 
the number of independent PDFs must be adjusted to match the partons to 
which the data is sensitive. 
Any linear combination of independent PDFs can be
used, for example the sum and difference of the quark and anti-quark 
distributions, ${\rm q}^+ \equiv {\rm q} + \bar{\rm q}$ and 
${\rm q}^- \equiv {\rm q} - \bar{\rm q}$, or the singlet and non-singlet 
distributions of the DGLAP evolution basis~\eqref{eq:DGLAP}.

\subsection{Optimization}
\label{subsec:optimization}

Once the parameterization is set at an input scale $Q_0$, the PDFs can be 
evolved to the scale of the data $Q$ by means of the DGLAP equations~\ref{eq:DGLAP},
and then convoluted with partonic cross sections to obtain theoretical 
predictions for the hadronic cross sections at a given perturbative order.
The optimal PDF parameters are then found by optimizing a suitable figure of 
merit, commonly taken as the log-likelihood $\chi^2$. Given a set of 
$N_{\rm dat}$ measurements $D_i$, and a set of corresponding theoretical 
predictions $T_i(\{ {\bm a}\})$, the $\chi^2$ is defined as
%-------------------------------------------------------------------------------
\begin{equation}
\chi^2({\bm a})
= 
\sum_{i,j}^{N_{\rm dat}} 
\left[D_i-T_i({\bm a})\right] 
{\rm cov}_{ij}^{-1} 
\left[D_j-T_j({\bm a})\right]
\,\mbox{,}
\label{eq:chi2cov}
\end{equation}
%-------------------------------------------------------------------------------
where the covariance matrix ${\rm cov}_{ij}$ is composed of all of the 
uncertainties of the data,
%-------------------------------------------------------------------------------
\begin{equation}
{\rm cov}_{ij}
=
\delta_{ij}\sigma_i^{\rm unc}\sigma_j^{\rm unc}
+
\sum_{k=1}^{n_{\rm corr}}\sigma_{k,i}^{\rm corr}\sigma_{k,j}^{\rm corr}
\,\mbox{,}
\label{eq:covariance}
\end{equation}
%-------------------------------------------------------------------------------
namely, uncorrelated ($\sigma_i^{\rm unc}$) and $n_{\rm corr}$ correlated 
($\sigma_{k,i}^{\rm corr,A}$) uncertainties. The $\chi^2$ also reads
%-------------------------------------------------------------------------------
\begin{equation}
\chi^2({\bm a})
= 
\sum_i^{N_{\rm dat}} \left(\frac{D_i 
+ 
\sum_j^{n_{\rm corr}} r_j \sigma_j^{\rm corr} - T_i(\{ {\bm a} \})}{\sigma_i^{\rm unc}}\right)^2
+ 
\sum_j^{n_{\rm corr}} r_j^2
\,\mbox{,}
\label{eq:chi2penalty}
\end{equation}
%-------------------------------------------------------------------------------
which is equivalent to Eq.~\eqref{eq:chi2cov} upon minimization with respect to 
the shift parameters, $r_j$. This allows one to study also the 
behavior of the shift parameters at the minimum, where
their distribution ought to be a univariate Gaussian with mean zero.

If correlated multiplicative uncertainties are provided ({\it e.g.} 
in the case of normalization errors such as the luminosity uncertainty), 
they must be treated carefully in order to avoid the so-called 
d'Agostini bias~\cite{DAgostini:1993arp}, which results in the optimal 
parameters giving a theoretical prediction that underestimates the data. 
To avoid this, multiplicative uncertainties in Eq.~\ref{eq:covariance}
can be treated iteratively, in which they are multiplied by central theory 
predictions from a previous fit instead of by the experimental data. This
procedure, called the $t_0$-method~\cite{Ball:2009qv}, was proven to rapidly
converge. Alternatively, one can fit an overall normalization parameter 
and allow it to fluctuate within the multiplicative uncertainty range. 
This can be naturally included in Eq.~\ref{eq:chi2penalty} as one of the $r_j$ 
parameters. 

The optimization of Eq.~\eqref{eq:chi2cov} is performed numerically, 
which means DGLAP evolution and convolutions between PDFs and 
partonic cross sections must be performed quickly, accurately, and precisely
in order to make global fits viable. 
In addition, fitting strategies must be optimized to explore 
a sufficiently large region of PDF parameter space.

Several publicly available codes solve DGLAP evolution 
equations efficiently~\cite{Vogt:2004ns,*Salam:2008qg,*Botje:2010ay,
Bertone:2013vaa,Bertone:2015cwa}. These programs, as well as 
most private evolution codes used in PDF fits, have been benchmarked 
against standard tables~\cite{Giele:2002hx,*Dittmar:2005ed,Bertone:2015cwa}. 
The calculation of hadronic cross sections
is performed via the use of lookup tables, where partonic cross sections
convoluted with evolution kernels are precomputed and stored for 
each data set on a suitable interpolation grid. Hadronic cross sections 
then reduce to the scalar product
between such interpolation tables and the PDFs parameterized at the scale $Q_0$.
This method is usually realized as part of each private optimization code
(see {\it e.g.}~\cite{Stratmann:2001pb}), except for the {\sc APFELgrid} 
program~\cite{Bertone:2016lga}. In this case, DGLAP evolution kernels and DIS 
partonic cross sections are provided by the {\sc APFEL} 
package~\cite{Bertone:2013vaa}, while partonic cross sections for other 
unpolarized processes are provided in the format of
{\sc APPLgrid}~\cite{Carli:2010rw} and {\sc FastNLO}~\cite{Wobisch:2011ij} 
tables. The latter are obtained in turn from multi-purpose Monte Carlo 
generators~\cite{Campbell:2019dru,*Nagy:2003tz,*Alwall:2014hca}, 
which are accurate to a given perturbative order and contain appropriate numerical 
interfaces~\cite{Bertone:2014zva,*DelDebbio:2013kxa}.

The choice of an efficient optimization strategy largely depends on the
dimension of the PDF parameter space. Numerical gradient-based algorithms 
such as the Newton's method, the Levenberg-Marquardt method (which supplements 
the Newton's method with the steepest descent method), or variable metric methods
(which only rely on gradient information) are used for spaces of 
moderate dimension (roughly on the order of 40 free parameters). If the 
dimension of the PDF parameter space is larger, these methods start to become
unsuitable due to inefficient numerical inversions of large matrices 
and the increased possibility of ending in a local minima. 
Stochastic genetic algorithms~\cite{Ball:2014uwa} or deterministic 
gradient descent methods, possibly associated with deep-learning and 
hyperoptimization techniques~\cite{Bertone:2017tyb,*Carrazza:2019mzf}, 
can be used to efficiently explore the parameter space in this case.
To avoid fitting the noise in the data, it is important to devise a suitable
stopping criterion. In this respect, a widely used method in the literature 
is cross-validation, where
the data points are randomly divided into two sets: training and validation. 
The $\chi^2$ is then computed for both sets separately, but optimized only on 
the training set. The fit terminates when the $\chi^2$ of the validation
set starts to increase (while the $\chi^2$ of the training set continues to 
decrease). To avoid information loss, the procedure should be repeated a 
sufficiently large number of times starting from different data partitions.

\subsection{Representation of PDF uncertainties}
\label{subsec:uncertainties}

The confidence interval in the space of PDFs, namely a representation of the 
PDF probability density, is intrinsically derived from 
the probability density of the fitted parameters given the data.
In Bayesian language we are interested in $\mathcal{P}({\bm a}|D)$, in which
the expectation $E$ and variance $V$ of an observable 
${\cal O}$ depending on a set of PDFs can be computed as,
%-------------------------------------------------------------------------------
\begin{equation}
E[{\cal O}] 
= 
\int d^n a\ {\cal P}({\bm a} | D)\ {\cal O}({\bm a})
\,\mbox{,}
\qquad
V[{\cal O}] 
= \int d^n a\ {\cal P}({\bm a} | D)\ ({\cal O}({\bm a}) - E[{\cal O}])^2
\,\mbox{.}
\label{e.EandV}
\end{equation}
%-------------------------------------------------------------------------------
In this section, we review the statistical estimation of $E$ and $V$ 
with focus on the three components that contribute to $V$:
experimental, procedural, and theoretical uncertainties.

\vspace{-0.3cm}
\paragraph*{Experimental uncertainties.}
There exists at least two commonly used approaches for propagating the data 
uncertainties into the PDFs: the Hessian and Monte Carlo methods.
 
The Hessian technique~\cite{Pumplin:2001ct} (including its Lagrange
multiplier generalization~\cite{Stump:2001gu}), adopted when a fairly 
limited parameterization is used, is based on the standard least-squares 
approach, and assumes that ${\cal P}({\bm a} | D)$ is multi-Gaussian. Once the 
best-fit is determined, the $\chi^2$ is approximated to first nontrivial 
order about its minimum. The desired confidence level is obtained as the volume
in parameter space about the minimum that corresponds to a fixed increase in 
$\chi^2$. For Gaussian uncertainties, the 68\% (or one-sigma) confidence level 
corresponds to the volume enclosed by $\Delta\chi^2=\chi^2-\chi^2_{\rm min}=1$. 
Given the set of parameters ${\bm a}_0$ that minimize the $\chi^2$ and the
Hessian matrix $H$ in the parameter space, whose elements are 
$H_{ij}=\frac{1}{2}\left.\frac{\partial^2\chi^2({\bm a})}{\partial a_i\partial a_j} \right|_{{\bm a}={\bm a}_0}$, it follows that
%-------------------------------------------------------------------------------
\begin{equation}
E[\mathcal{O}]
=
\mathcal{O}({\bm a}_0)
\,\mbox{,}
\qquad
V[\mathcal{O}]
=
\sum_{i,j}^{n_{\rm par}}\left.\frac{\partial\mathcal{O}}{\partial a_1}H_{ij}^{-1}\frac{\partial\mathcal{O}}{\partial a_j}\right|_{{\bm a}={\bm a}_0}
=
\sum_k^{n_{\rm eig}}(\Delta\mathcal{O}_k)^2
\,\mbox{.}
\label{eq:Hessian}
\end{equation}
%-------------------------------------------------------------------------------
Here, $n_{\rm par}$ is the number of free parameters and $n_{\rm eig}$ is the number
of eigenvectors of the matrix $H$. The confidence level is therefore entirely 
determined by the inverse of the Hessian matrix, which is equivalent to the 
covariance matrix in parameter space. Uncertainties on the PDFs, and on any 
quantities depending on them, can be compactly represented by providing
eigenvectors of the Hessian matrix rescaled by their eigenvalues. They
can then be computed by adding in quadrature the 
variation along the direction of each eigenvector by a fixed amount,
$\Delta\mathcal{O}_k=\mathcal{O}({\bm a}_i)-\mathcal({\bm a}_0)$
(more sophisticated formulas hold for asymmetric eigenvectors).
However, data sets can be incompatible,
in which case the prescription $\Delta\chi^2=1$ for
the parameter shifts may lead to 
unrealistically small uncertainties. To compensate this, the scaling along the 
eigendirections is inflated by a tolerance factor $T=\sqrt{\Delta\chi^2}>1$. 
Its specific value is usually chosen by studying the distribution of best-fit 
parameters across experiments or by determining a different tolerance along 
each Hessian eigenvector~\cite{Martin:2009iq}. 

In addition to the tolerance criteria, there are two additional disadvantages
to the Hessian method. The first is that it becomes rapidly unviable 
when the parameter space is large. Secondly,
the Gaussian assumption, and also a linear approximation for the
observable ${\cal O}$ that is made in deriving Eq.~\ref{eq:Hessian},
might become inadequate to handle flat directions that arise whenever
the data loosely constrain the PDF parameters.

A more robust method is the Monte Carlo technique, which samples 
the probability distribution ${\cal P}({\bm a} | D)$ by bootstrapping 
the starting data sample ({\it i.e.} smearing the experimental data 
about their respective uncertainties) and performing a fit to each
of these new data replicas. 
The distribution of the data is therefore mapped into the distribution of the 
PDFs, provided that the number of data replicas $n_{\rm rep}$ is sufficiently 
large (typically on the order of a few hundred). The expectation value and 
uncertainty of any quantity depending on the PDFs can therefore be 
computed as a mean and variance over the PDF ensemble,
%-------------------------------------------------------------------------------
\begin{equation}
E[{\cal O}] 
=
\frac{1}{n_{\rm rep}} \sum_k^{n_{\rm rep}} {\cal O}({\bm a}_{0,k})
\,\mbox{,}
\qquad
V[{\cal O}] 
=
\frac{1}{n_{\rm rep}} \sum_k^{n_{\rm rep}} ({\cal O}({\bm a}_{0,k}) - E[{\cal O}])^2
\,\mbox{,}
\label{eq:MonteCarlo}
\end{equation}
%-------------------------------------------------------------------------------
where ${\bm a}_{0,k}$ are the best-fit parameters for each replica. 

Despite being computationally expensive, the Monte Carlo method has some 
obvious advantages with respect to the Hessian method:
any distribution can be used to smear the data, should they be non-Gaussian;
there is no need to rely on the linear approximation of observables; there are 
no limitations in how large the parameter space can be; and the 
probability density ${\cal P}({\bm a} | D)$ can be readily updated with 
Bayes' theorem to incorporate a new piece of experimental information without 
performing a new fit. The last of these is realized through 
reweighting~\cite{Ball:2010gb,*Ball:2011gg}, which consists of assigning each
PDF replica a weight that quantifies its agreement with the new data.
In this case, formulas in Eq.~\eqref{eq:MonteCarlo} should be replaced 
by their weighted counterparts. Since replicas with 
small weights become immaterial, the PDF
ensemble looses part of its statistical power, a drawback of  
the method when such a loss is excessive.
In any case, Monte Carlo uncertainties are 
statistically sound in that they can be rigorously validated (the same is not
true for the tolerance criterion). This can be done by performing a closure 
test~\cite{Ball:2014uwa}, where one assumes that the underlying parton
distributions are known by using a specific (previously determined) 
PDF set to generate artificial data. Fits are
then performed to check whether the output PDFs provide 
consistent and unbiased estimators of the underlying truth, and that
the confidence levels reproduce the correct coverage, among other
important considerations.

Techniques have also been developed to convert Hessian PDF sets to Monte Carlo 
PDF sets and vice versa~\cite{Hou:2016sho,*Carrazza:2015aoa}, and to optimize
the number of replicas in a PDF ensemble without statistical 
loss~\cite{Carrazza:2015hva}. Such methods have led to an 
extension of the reweighting procedure to Hessian 
sets~\cite{Paukkunen:2014zia}, 
the construction of statistical combinations of different PDF 
sets~\cite{Butterworth:2015oua}, to optimized versions of
PDF sets~\cite{Schmidt:2018hvu,Carrazza:2016htc}, and to statistical tools 
that analyze the sensitivity of the data set to PDFs (and to any quantity 
depending on them~\cite{Wang:2018heo}. Finally, we shall note that additional 
optimization algorithms, not based
on least-squares, exist to determine the probability density 
${\cal P}({\bm a} | D)$ of the PDF parameters. Among these are 
Nested Sampling and Markov chain Monte Carlo methods, which
have been pioneered in~\cite{Lin:2017stx,*Gbedo:2017eyp}.

\vspace{-0.3cm}
\paragraph*{Procedural uncertainties.} 
There are three classes of procedural uncertainties, all connected to 
optimization and naturally incorporated in ${\cal P}({\bm a} | D)$. 
The first is the methodological uncertainty related to the 
choices made in a fit, namely the basis functions, the functional form, the 
number of parameters, and the minimization strategy. The second is the 
extrapolation uncertainty related to the fact that data points, even when 
infinitely precise, are not covering the entire phase space. The third is the 
functional uncertainty related to the fact that a set of functions (which are 
infinite dimensional objects) is determined from a finite amount of 
information. Within the Hessian method, the three classes of uncertainties
are usually assessed by comparing different fits obtained with varied
procedural input, and accounted for by the tolerance criterion.
In the Monte Carlo method, the methodological uncertainty can be
made immaterial by tuning the fitting procedure. 
Such a tuning is achieved
through a level-0 closure test, in which perfect data (no uncertainties) 
are generated from an assumed underlying law. The test is successful if a 
perfect fit to the data can be produced, {\it i.e.} a fit with vanishing 
$\chi^2$. The extrapolation and functional uncertainties are irreducible, 
yet they can be completely characterized. The former can be 
determined by looking at the remaining uncertainty on the final PDF in a 
level-0 closure test. The latter can be determined in a level-1 closure test, 
where the data replicas are fluctuated to mimic statistical noise 
(but not the uncertainty of the data itself). The 
functional uncertainty then arises from all of the statistically equivalent fits. 
In QCD analyses, closure tests were realized completely for unpolarized 
PDFs~\cite{Ball:2014uwa} and for nuclear PDFs~\cite{AbdulKhalek:2019mzd}.

\vspace{-0.3cm}
\paragraph*{Theoretical uncertainties.}
There is a variety of theoretical uncertainties related to the assumptions 
made in the computation of hadronic observables that are more challenging to 
propagate into PDF uncertainties. These include missing higher-order uncertainties (MHOU) 
due to the truncation of the perturbative expansion to a given order,
uncertainties in the input values of the physical parameters such as
the strong coupling and the heavy quark masses, uncertainties due to the 
neglect of power-suppressed corrections should they not be included in 
factorization formulas, and uncertainties due to nuclear corrections when data 
on nuclear targets are used in fits of proton PDFs. The first is usually 
estimated via scale variations, see Eq.~\eqref{eq:MIssHO}. The
second is accounted for by performing different fits with 
varied values of the input physical parameters and combining the results. 
In this respect, the Hessian method allows these uncertainties to be treated as 
additional sources of nuisance by summing them in quadrature in Eq.~\eqref{eq:Hessian}. 
The third can usually be kept under control by removing data points
from the fit that are particularly sensitive to power
suppressed corrections and
by looking at the stability of the fit upon variations of this  
cut-off~\cite{Martin:2003sk,Accardi:2009br}. Alternatively, power corrections 
can be modeled and fitted along with PDFs~\cite{Accardi:2016qay,
*Sato:2016tuz}, as well as nuclear corrections~\cite{Accardi:2016qay,
*Sato:2016tuz,Martin:2009iq,Ball:2009mk,*Martin:2012da}, and uncertainties 
estimated from PDF variations.

A general procedure to represent theory uncertainties in PDFs has been
proposed recently in the framework of the Monte Carlo
method~\cite{Ball:2018odr}. 
Assuming that they are Gaussian, it follows from Bayes' theorem that 
they can be included by redefining the covariance matrix in
Eq.~\eqref{eq:chi2cov} as the sum of an experimental and a theoretical part,
${\rm cov}_{ij}={\rm cov}_{ij}^{\rm exp}+{\rm cov}_{ij}^{\rm th}$. The resulting 
$\chi^2$ is then used both in the sampling of the data replicas and in the
optimization of the fit. The problem of propagating theory uncertainties into
PDFs is therefore reduced to estimating the theoretical covariance matrix 
${\rm cov}_{ij}^{\rm th}$ by way of an educated guess. Estimates were 
formulated so far in two cases of unpolarized PDFs studies. First, for nuclear 
uncertainties by defining the matrix elements of ${\rm cov}_{ij}^{\rm th}$ as the
difference between theoretical predictions obtained either with a free proton 
or a nuclear PDF, and then taking an average over replicas~\cite{Ball:2018twp}. 
Secondly, for MHOU (at NLO) by defining the matrix elements of 
${\rm cov}_{ij}^{\rm th}$ as the difference between theoretical predictions 
obtained with either central or varied factorization and renormalization scales
according to various prescriptions~\cite{AbdulKhalek:2019bux,
*AbdulKhalek:2019ihb}. In these studies, correlations across data points
induced either by the nuclear target in the first case,
or by the structure of higher-order corrections in the partonic cross sections
and splitting functions in the second, were accounted for properly.
Furthermore, nuclear uncertainties and MHOU were validated
against the exact nuclear and NNLO results, respectively.
The inclusion of such theoretical uncertainties improves the description of
the data, shifts the central value of the PDFs towards the truth, and 
slightly increases their uncertainties.

\section{STATE-OF-THE-ART RESULTS}
\label{sec:results}

As is evident by the previous sections,
the determination of unpolarized, polarized, and nuclear PDFs
is a particularly involved problem. It is therefore addressed 
by various collaborations of physicists who 
regularly produce and update general-purpose PDF sets, many of which
have a history as long as two decades (see Sect.~1 in~\cite{Forte:2013wc} 
and Sect.~2.1 in~\cite{Gao:2017yyd} for an overview). Furthermore,
while most collaborations
perform their global QCD analyses privately, the xFitter
collaboration has developed an open-source fitting 
framework~\cite{Alekhin:2014irh}. 
Most of the recent PDF determinations are summarized with their
theoretical, experimental, and methodological features in 
Table~\ref{tab:pdfs_all}. All but LSS15, DSSV14 and JAM17, are
publicly available through the LHAPDF library~\cite{Buckley:2014ana} and
can be readily visualized on-line~\cite{Carrazza:2014gfa}.
In this section, we delineate the current status of 
unpolarized, polarized, and nuclear PDFs using recent PDF extractions.

%------------------------------------------------------------------------------
\begin{table}[!p]
\centering
\begin{tabular}{lcccc}
\toprule
  \sc Unpolarised PDFs 
& CT18~\cite{Hou:2019efy}
& MMHT14~\cite{Harland-Lang:2014zoa}
& NNPDF3.1~\cite{Ball:2017nwa} 
& JAM19~\cite{Sato:2019yez} 
\\
\midrule
  Perturbative order      
& NLO, NNLO  
& LO, NLO, NNLO 
& LO, NLO, NNLO 
& NLO 
\\ 
  Heavy quark scheme      
& S-ACOT     
& optimal-TR    
& FONLL   
& ZM-VFN    
\\
  Value of $\alpha_s(m_Z)$ 
& 0.118      
& 0.118         
& 0.118   
& 0.118     
\\
  Input scale $Q_0$
& 1.30 GeV
& 1.00 GeV
& 1.65 GeV
& 1.27 GeV
\\
\midrule
  Fixed Target DIS    
& \checkmark 
& \checkmark 
& \checkmark 
& \checkmark 
\\
  Collider DIS        
& \checkmark 
& \checkmark 
& \checkmark 
& \checkmark 
\\
  Fixed Target SIDIS  
&            
&            
&            
& \checkmark 
\\
  Fixed Target DY     
& \checkmark 
& \checkmark 
& \checkmark 
& \checkmark 
\\
  Collider DY         
& \checkmark 
& \checkmark 
& \checkmark 
&            
\\
  Jet production      
& \checkmark 
& \checkmark 
& \checkmark 
&            
\\
  Top production      
& $t\bar{t}$ tot., diff.
& $t\bar{t}$ tot.
& $t\bar{t}$ tot., diff.
&            
\\
  \midrule  
  Independent PDFs 
& 6
& 7
& 8
& 7
\\
  Parameterization
& Bernstein pol. 
& Chebyshev pol.
& neural network 
& simple pol.
\\
  Free parameters
& 29
& 37
& 296
& 19
\\
  Statistical treatment      
& Hessian
& Hessian
& Monte Carlo
& Monte Carlo 
\\
  Tolerance
& $\Delta\chi^2=100$
& $\Delta\chi^2$ dynamical
& n/a
& n/a         
\\
\midrule
{\sc Polarised PDFs} 
& LSS15~\cite{Leader:2014uua}
& DSSV14~\cite{deFlorian:2019zkl} 
& NNPDFp1.1~\cite{Nocera:2014gqa}
& JAM17~\cite{Ethier:2017zbq} 
\\
\midrule
  Perturbative order
& NLO
& NLO
& NLO
& NLO
\\
  Heavy quark scheme
& ZM-VFN
& ZM-VFN
& ZM-VFN
& ZM-VFN
\\
  Value of $\alpha_s(m_Z)$
& 0.118
& 0.120
& 0.118
& 0.118
\\
Input scale $Q_0$
& 1.00 GeV
& 1.00 GeV
& 1.00 GeV
& 1.00 GeV
\\
\midrule
  Fixed Target DIS
& \checkmark
& \checkmark
& \checkmark
& \checkmark
\\
  Fixed Target SIDIS
&
& \checkmark
&
& \checkmark
\\
  Colider DY
&
&
& \checkmark
&
\\
  Jet and had. prod.
& 
& \checkmark
& (jet only)
&
\\
\midrule
  Independent PDFs
& 4
& 6
& 6
& 6
\\
Parameterization
& simple pol.
& simple pol.
& neural network
& simple pol.
\\
Free parameters
& 13
& 19
& 148
& 24
\\
Statistical treatment
& Hessian
& Monte Carlo
& Monte Carlo
& Monte Carlo
\\
\midrule
{\sc Nuclear PDFs} 
& nCTEQ15~\cite{Kovarik:2015cma}
& EPPS16~\cite{Eskola:2016oht}
& nNNPDF1.0~\cite{AbdulKhalek:2019mzd}
& TUJU19~\cite{Walt:2019slu} 
\\
\midrule
  Perturbative order
& NLO
& NLO
& NLO, NNLO
& NLO, NNLO
\\
  Heavy quark scheme
& ACOT
& S-ACOT
& FONLL
& ZM-VFN
\\
  Value of $\alpha_s(m_Z)$
& 0.118
& 0.118
& 0.118
& 0.118
\\
  Input scale $Q_0$
& 1.30 GeV
& 1.30 GeV
& 1.00 GeV
& 1.30 GeV
\\
\midrule
  Fixed Target DIS
& \checkmark
& \checkmark
& (w/o $\nu$-DIS)
& \checkmark
\\
  Fixed Target DY
& \checkmark
& \checkmark
&
&
\\
  Colider DY
&
& \checkmark
& 
&
\\
  Jet and had. prod.
& ($\pi^0$ only)
& ($\pi^0$, dijet)
& 
&
\\
\midrule
  Independent PDFs
& 6
& 6
& 3
& 6
\\
  Parameterization
& simple pol.
& simple pol.
& neural network
& simple pol.
\\
  Free parameters
& 16
& 20
& 178
& 16 
\\
  Statistical treatment
& Hessian
& Hessian
& Monte Carlo
& Hessian
\\
  Tolerance
& $\Delta\chi^2=35$
& $\Delta\chi^2=52$
& n/a
& $\Delta\chi^2=50$
\\
\bottomrule
\end{tabular}

\caption{A summary of the theoretical, experimental, and methodological features
(see Sects.~\ref{sec:theory}-\ref{sec:methodology}) for the most
recent unpolarized, polarized, and nuclear PDF sets.}
\label{tab:pdfs_all}
\end{table}
%-------------------------------------------------------------------------------

%-------------------------------------------------------------------------------
\begin{figure}[!t]
\centering
\includegraphics[width=\textwidth]{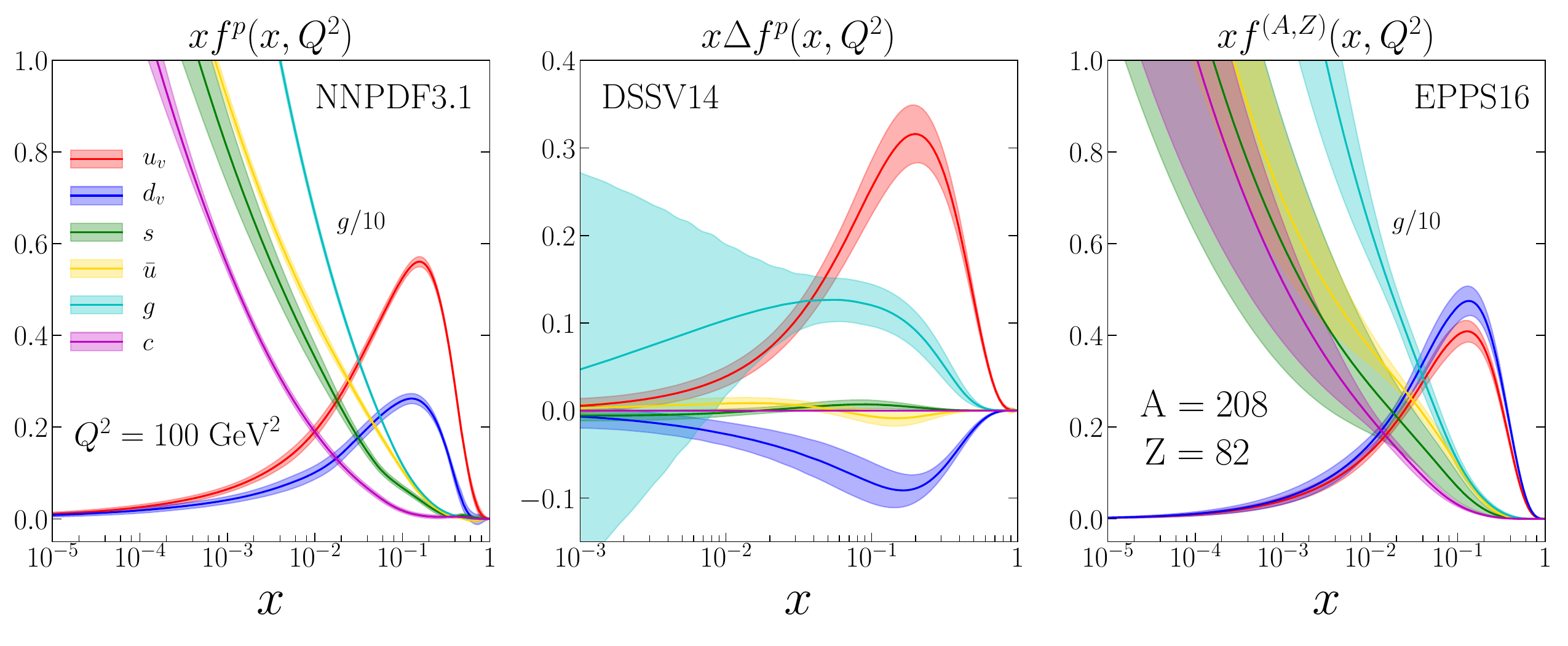}\\
\caption{A representative snapshot of unpolarized and polarized 
proton PDFs, and on nuclear (lead) PDFs. Parton distributions are taken
from the NNPDF3.1 (NNLO), DSSV14 (NLO) and EPPS16 (NLO) analyses, respectively. 
Uncertainty bands correspond to Monte Carlo 68\% confidence levels for NNPDF3.1 
and DSSV14, and to Hessian 90\% confidence levels for EPPS16.}
\label{fig:pdfs_all}
\end{figure}
%-------------------------------------------------------------------------------

\subsection{Unpolarized PDFs}
\label{subsec:unpolarized}

The most recent global determinations of unpolarized PDFs are 
CT18~\cite{Hou:2019efy}, MMHT14~\cite{Harland-Lang:2014zoa}, 
NNPDF3.1~\cite{Ball:2017nwa}, JAM19~\cite{Sato:2019yez}
and ABMP16~\cite{Alekhin:2017kpj,*Alekhin:2018pai}.
Since their publication, the MMHT14 and the NNPDF3.1 analyses have been updated
with new data and improved theoretical frameworks. In particular, the first was extended to
include the HERA I-II legacy measurements~\cite{Harland-Lang:2016yfn},
as well as differential measurements in top-pair production~\cite{Bailey:2019yze} and 
recent jet production measurements~\cite{Harland-Lang:2017ytb} from the LHC.
It was also extended to a simultaneous determination of 
$\alpha_s$~\cite{Harland-Lang:2015nxa} and of the photon 
PDF~\cite{Harland-Lang:2019pla}. The NNPDF3.1 analysis has since included
direct photon~\cite{Campbell:2018wfu} and single-top 
production~\cite{Nocera:2019wyk} measurements at the LHC. Similar to MMHT14,
it also extended to a determination of
$\alpha_s$~\cite{Ball:2018iqk} and of the photon PDF~\cite{Bertone:2017bme}.
Lastly, the NNPDF collaboration recently assessed theoretical uncertainties from 
nuclear corrections~\cite{Ball:2018twp} and missing higher 
orders~\cite{AbdulKhalek:2019bux,*AbdulKhalek:2019ihb},
as well as studied the impact of small-$x$ resummation~\cite{Ball:2017otu}.

The differences among these PDF sets are summarized in Table~\ref{tab:pdfs_all} 
except for ABMP16, which is the only unpolarized PDF set determined in the FFN 
scheme for $n_f=3,4,5$ active flavors. In this case, $\alpha_s$ and the heavy quark 
masses were also free parameters determined together with the PDFs, 
and each of the six PDF flavors 
were parameterized in terms of a simple polynomial 
for a total of 25 free parameters. In addition, ABMP16 
utilized the Hessian method for error propagation, 
with $\Delta\chi^2=1$. All of the different PDF
sets are based on a fairly similar dataset 
and are available at NLO and NNLO except for JAM19, which is the only PDF 
set that achieved a simultaneous NLO determination of unpolarized proton 
PDFs and FFs from charged hadron production in SIDIS and electron-positron 
annihilation. Additional unpolarized PDF sets exist in the literature, namely 
JR14~\cite{Jimenez-Delgado:2014twa}, CJ15~\cite{Accardi:2016qay} and
HERAPDF2.0~\cite{Abramowicz:2015mha}, although these PDF sets are based 
on a reduced set of measurements and are somewhat more limited in scope. 
In particular, HERAPDF2.0 was an 
analysis of only the HERA I-II legacy data, JR14 studied the impact of
a valence-like input below $Q^2\sim 1$ GeV$^2$ (see also~\cite{Diehl:2019fsz}),
and CJ15 focused on the high-$x$ data region 
via the assessment of power-suppressed corrections.

%-------------------------------------------------------------------------------
\begin{figure}[!t]
\centering
\includegraphics[width=\textwidth]{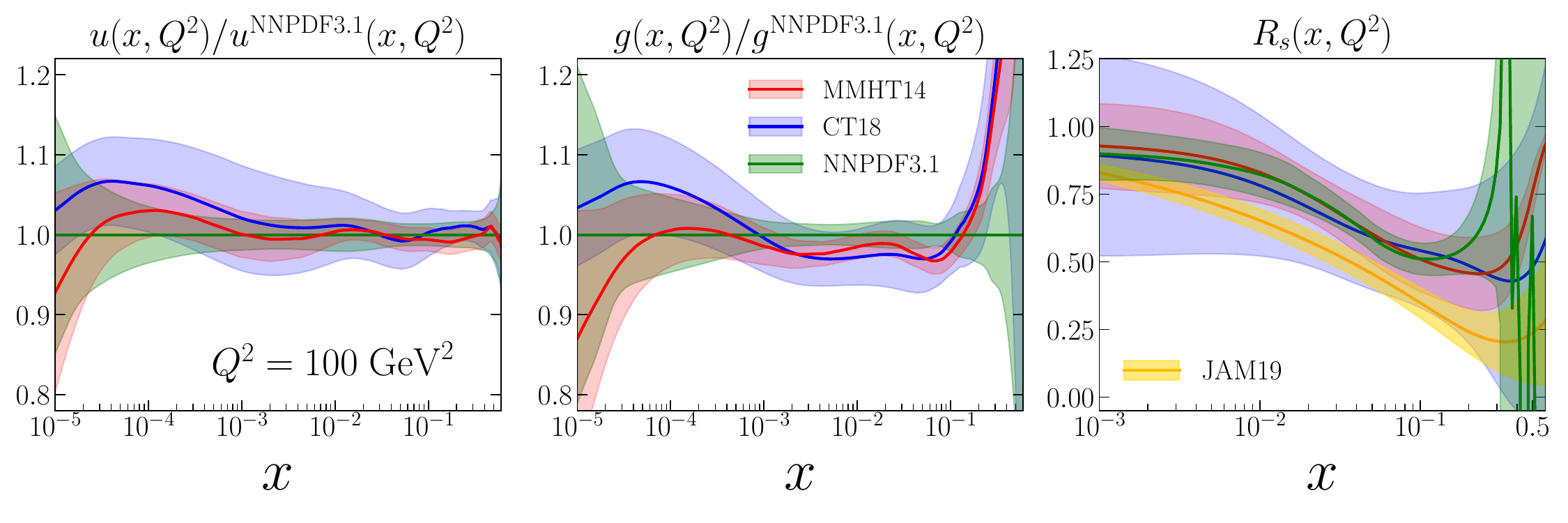}\\
\caption{Representative examples of unpolarized PDFs as a function of $x$ at 
$Q^2=100$ GeV$^2$ for the up quark (left) and gluon (middle) PDFs normalized to the
NNPDF3.1 central value, and for the ratio
$R_s=\frac{s+\bar{s}}{\bar{u}+\bar{d}}$ (right). Results correspond to NNLO 
MMHT14, CT18, NNPDF3.1 and NLO JAM19 PDF sets. Uncertainty bands 
computed as 68\% confidence levels.}
\label{fig:upol_comparison}
\end{figure}
%-------------------------------------------------------------------------------

Representative unpolarized PDFs from NNPDF3.1 are presented
as a function of $x$ at $Q^2=100$ GeV$^2$ 
in the left panel of Fig.~\ref{fig:pdfs_all} for 
all flavors, and in Fig.~\ref{fig:upol_comparison} for the up quark and gluon 
distributions normalized to the NNPDF3.1 result. Also illustrated in Fig.~\ref{fig:upol_comparison} 
is the sea quark ratio
$R_s=(s+\bar{s})/(\bar{u}+\bar{d})$. Results correspond to the NNLO PDF sets 
except for JAM, which is displayed at NLO, and all uncertainty bands correspond to 
68\% confidence levels. The general features of unpolarized PDFs can be 
summarized as follows.

\vspace{-0.3cm}
\paragraph*{Valence quarks.}
The relevance of quark valence distributions is twofold. First,
they give the global properties of the 
nucleon across their entire range in $x$, 
such as its charge and baryon number. Furthermore, at large
$x$ they carry most of the nucleon's momentum and thus are
sensitive to its non-perturbative structure and to the production of new
particles, {\it e.g.} heavy $W^\prime$ and $Z^\prime$ bosons at high rapidities 
and invariant masses. Up and down valence distributions are primarily constrained by
collider DY measurements at small $x$ (in particular by the $W^\pm$
asymmetry), by neutral- and charged-current DIS measurements at medium $x$, 
and by fixed-target DY measurements at medium to high $x$.
Results obtained from various PDF sets agree very well for the up 
valence distribution, which has an overall relative uncertainty of a few percent 
in the data region, while they are a little more widespread for the down
valence distribution (see Sect.~6.2 in~\cite{Gao:2017yyd} for more details).

\vspace{-0.3cm}
\paragraph*{Sea quarks.}
Anti-up and anti-down sea quarks are probed by parity-violating processes,
in particular by charged-current DIS and DY measurements. Unlike the 
valence distributions, they are strongly suppressed at large $x$ and show a 
steep rise at small $x$ due to being produced mainly
by gluon splitting. Anti-up and anti-down PDFs display 
larger uncertainties than their valence 
distribution counterparts, with more marked differences across PDF sets.

Concerning the strange quark PDF, discrepancies have arisen in analyses of 
inclusive $W^\pm$ production from ATLAS with respect to charged current 
neutrino DIS measurements. While the former support a ratio of the strange to 
non-strange light sea distributions around unity, $R_s \sim 1.13$ at $x=0.023$ 
and scale $Q^2 = 1.9$ GeV$^2$, the latter give a result $R_s \sim 0.5$ at similar 
kinematics~\cite{Ball:2017nwa}. The tension may be somewhat relieved if a 
flexible parameterization is used~\cite{Ball:2017nwa}, if massive corrections
to neutrino DIS structure functions are included in the analysis, or if 
experimental correlations are relaxed~\cite{Kassabov:2019xxx}. 
The picture is further complicated by recent analyses of charged kaon 
production in SIDIS, which can provide crucial strange quark PDF information 
due to the enhancement from the favored $s(\bar{s})\rightarrow K^- (K^+)$ FF.
Using a reweighting procedure, a minor suppression of $R_s$ was found
in the intermediate $x$ region compared to the result obtained without 
the inclusion of SIDIS data~\cite{Borsa:2017vwy}. On the other hand, 
the JAM19 analysis found a more suppressed $R_s$
result by simultaneously fitting the FFs with the proton PDFs~\cite{Sato:2019yez}. 
All of the various $R_s$ distributions can be seen in Fig.~\ref{fig:upol_comparison}, 
where uncertainties are rather large. To alleviate this,  
better understanding of the various systematic 
uncertainties that affect the data (especially nuclear modifications in 
neutrino-nucleus DIS) and clean processes that are sensitive to strangeness
are desperately needed.
One such process is $W$-boson production in association with a charm quark.
However, in this case important NNLO corrections to the theoretical prediction
remain absent.

\vspace{-0.3cm}
\paragraph*{Heavy quarks.}
Heavy quarks produce sizable contributions to inclusive DIS structure functions, typically
about 30\% (3\%) in $F_2$ for the charm (bottom). They are usually determined
perturbatively through gluon splitting in quark-antiquark pairs. Under this assumption,
the distributions turn out to be consistent across different PDF sets, with a precision 
similar to that of the gluon PDF (see below). An intrinsic, non-perturbative
heavy quark component can also be assumed, however. In this case, heavy quark PDFs
must be parameterized and determined together with the light quark PDFs.
This approach is default for the treatment of the charm in NNPDF3.1,
which leads to a reduced dependence on the charm mass for several high-energy
benchmark cross sections, to an overall improvement in the description of the 
data, and to a more stable gluon PDF (see also 
Sect.~\ref{susec:nonperturbative} for non-perturbative implications).

\vspace{-0.3cm}
\paragraph*{Gluon.}
The gluon PDF is probed in various ways. At small to medium $x$, it
can be accessed through 
measurements of the structure function $F_L$ in inclusive DIS as well as
in scaling violations in DGLAP evolution. At medium 
to large $x$, measurements of the $Z$-boson transverse momentum 
distributions can provide information. Lastly at large $x$, the gluon
can be constrained by jet and top-pair 
production measurements. The available measurements for the last
two classes of processes have recently raised some controversy, where
it was argued that they can be well described only if
experimental correlations are relaxed. For jets, such decorrelations 
do not affect the ensuing PDFs~\cite{Harland-Lang:2017ytb,Hou:2019efy}, 
while they might for top production~\cite{Bailey:2019yze,Hou:2019efy}, especially in 
relationship to the specific combination of kinematic distributions included 
in the fit. For this reason, the comparison of various PDF sets displays 
different shapes for the gluon PDF in Fig.~\ref{fig:upol_comparison}, 
particularly at large $x$, albeit with 
large uncertainties. This behavior may challenge the study of high-invariant 
mass states beyond the SM, possibly in conjunction with threshold 
resummation~\cite{Bonvini:2015ira,*Beenakker:2015rna}.
At very small $x$, the gluon remains largely unconstrained; however, 
small-$x$ resummation was recently demonstrated to point towards
evidence of BFKL dynamics~\cite{Ball:2017otu}.

\vspace{-0.3cm}
\paragraph*{Photon.}
It has been recently shown that the photon PDF, relevant in several electroweak
processes through photon-initiated and higher-order contributions, can be 
completely determined from the structure functions of the proton $F_2$ and
$F_L$~\cite{Manohar:2016nzj,*Manohar:2017eqh}. The computation, known as
LUXqed formalism, made previous approaches based on a model
assumption~\cite{Martin:2004dh,*Schmidt:2015zda} 
and a loosely constrained parameterization~\cite{Ball:2013hta} 
obsolete. Recent analyses~\cite{Harland-Lang:2019pla,Bertone:2017bme}
supplemented the LUXqed formalism with constraints from the data.
In these cases, the photon PDF resulted in an uncertainty of few percent, 
carrying about $0.5\%$ of the proton's momentum, and accounted for 
corrections of up to $20\%$ in DY, vector-boson, top quark and Higgs 
production processes at the LHC (see Sect.~7 in~\cite{Gao:2017yyd}
for additional details).

\subsection{Polarized PDFs}
\label{subsec:polarized}

The most recent analyses of polarized PDFs are 
LSS15~\cite{Leader:2014uua}, DSSV14~\cite{deFlorian:2014yva},
NNPDFpol1.1~\cite{Nocera:2014gqa} and JAM17~\cite{Ethier:2017zbq}.
Since publication, the DSSV14 analysis has been updated with a 
Monte Carlo variant~\cite{deFlorian:2019zkl}, which studied
the impact of recent di-jet measurements from 
STAR~\cite{Adamczyk:2016okk,*Adam:2018pns}. 
The NNPDFpol1.1 PDF set was also updated with 
STAR measurements, including the same 
di-jet measurements and additional 
$W$-boson production data~\cite{Nocera:2017wep}.

The similarities and differences among the polarized PDF sets are summarized in 
Table~\ref{tab:pdfs_all}. Here, all of the global QCD fits are performed
at NLO accuracy and are more or less consistent in their theoretical and methodological 
choices. However, each analysis supplements
inclusive DIS measurements with different data. 
The JAM17 and DSSV14 PDF extractions incorporated spin asymmetry measurements 
from pion and kaon production in SIDIS by HERMES~\cite{Airapetian:2004zf}
and COMPASS~\cite{Alekseev:2009ac,Alekseev:2010ub}. 
Single-jet production asymmetries in polarized 
proton-proton collisions measured by STAR~\cite{Adamczyk:2012qj,*Adamczyk:2014ozi} 
were used in the DSSV14 and NNPDFpol1.1 analyses. The DSSV14
dataset also included neutral pion production
in polarized proton-proton collision measurements 
by PHENIX~\cite{Adare:2008qb,*Adare:2008aa,*Adare:2014hsq}, while
the NNPDFpol1.1 data contained open-charm production asymmetries in DIS from 
COMPASS~\cite{Adolph:2012ca} and $W$-boson production asymmetries 
in polarized proton-proton collisions from STAR~\cite{Adamczyk:2014xyw}. 
Lastly, while not indicated in Table~\ref{tab:pdfs_all}, the JAM17 dataset
included charged pion and kaon production from 
single-inclusive electron-positron annihilation to help constrain
the final state FFs simultaneously with the polarized PDFs~\cite{Ethier:2017zbq}.

%-------------------------------------------------------------------------------
\begin{figure}[!t]
\centering
\includegraphics[width=\textwidth]{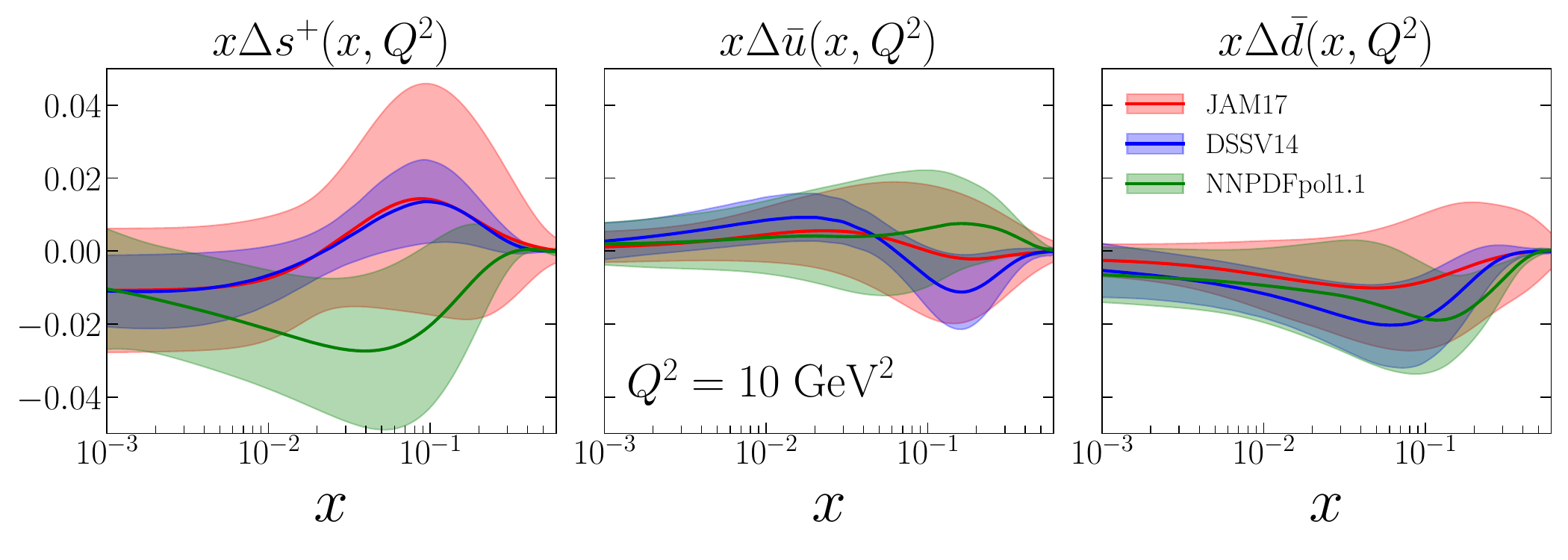}\\
\caption{Polarized PDFs as a function of $x$ 
for the total strange (left), anti-up (middle) and 
anti-down (right) PDFs at $Q^2=10$ GeV$^2$. 
Displayed are the JAM17 (red), DSSV14 (blue),
and NNPDFpol1.1 (green)
NLO PDF sets. Uncertainty bands are computed as 
68\% confidence levels.}
\label{fig:ppol_comparison}
\end{figure}
%-------------------------------------------------------------------------------

Representative examples of polarized PDFs at NLO are illustrated as a function of $x$ 
in the central panel of Fig.~\ref{fig:pdfs_all} for all flavors at 
$Q^2=100$~GeV$^2$, and in Fig.~\ref{fig:ppol_comparison} for the total strange
($x\Delta s^+=x(\Delta s +\Delta\bar{s})$), anti-up ($x\Delta\bar{u}$), 
and anti-down ($x\Delta\bar{d}$) distributions
at $Q^2=10$ GeV$^2$. In the latter figure, results are 
compared between the JAM17, DSSV14, and NNPDFpol1.1 PDF sets. 
Moreover, all uncertainty bands correspond to 68\% confidence levels.
The general features of polarized PDFs can be summarized as follows.

\vspace{-0.3cm}
\paragraph*{Valence quarks.}
The up quark valence distribution is the most constrained polarized PDF,
primarily due to measurements of the proton structure function $g_1$ over a relatively
broad range of $x$ and $Q^2$. The corresponding down quark distribution, which 
is opposite in sign, is smaller in magnitude and shows somewhat larger 
uncertainties. Nevertheless, a fair agreement between various PDF sets 
has been achieved for the valence polarizations, with differences 
originating from theoretical, experimental, and 
methodological choices generally being smaller than the PDF 
uncertainty, {\it i.e.} the uncertainty of the data.

\vspace{-0.3cm}
\paragraph*{Sea quarks.}
As can be seen in Fig.~\ref{fig:pdfs_all}, the polarization of the 
sea quark is significantly smaller than the 
polarization of the valence quarks. It is also more
dependent on the measurements included in each PDF set and on 
the flavor assumptions with which the data are analyzed. While all PDF sets show a
fairly consistent anti-down quark polarization, the anti-up distribution has opposite 
sign for the DSSV14 and NNPDFpol1.1 sets in the region 
$0.1\lesssim x\lesssim 0.5$. Recall, however, that this is driven by 
different processes in the two PDF sets, particularly by fixed-target 
SIDIS multiplicities and by collider $W$-boson production spin asymmetries, 
respectively. 
The former requires the knowledge of FFs~\eqref{eq:SIDISfact},
which are taken as a fixed input in the DSSV14 analysis. The difference is 
somewhat relieved by the JAM17 result, where the FFs entering the analysis of 
the SIDIS data were determined together with the polarized PDFs. 

The situation is even more involved in the case of the polarized strange PDF. 
Here, the distribution is entirely unconstrained in PDF sets that do not include SIDIS
asymmetries from the kaon sector unless one imposes the SU$_f$(3)
constraint from weak baryon decays (Eq.~\eqref{eq:gA}). This is the case,
for instance, in the NNPDFpol1.1 analysis, which displays a negative strange 
polarization peaked at $x\sim 0.1$ as a result. In contrast, the DSSV14 PDF set,
based on SIDIS data with a fixed kaon FF, finds a sign-changing
strange PDF that is incompatible with the NNPDFpol1.1 result in the range 
$0.02\lesssim x\lesssim 0.2$. This discrepancy is again somewhat alleviated 
within the large uncertainties of the JAM17 result, obtained by removing the 
SU$_f$(3) constraint from the analysis and by fitting the kaon FFs
simultaneously with the polarized PDFs. In any case, more precise 
SIDIS polarization asymmetries for
kaon production are desirable to reduce the uncertainty of the strange
PDF and shed light on possible SU$_f$(3) symmetry breaking effects. 
Moreover, separate measurements for
positively and negatively charged kaons could in principle allow for a
separation of the strange and anti-strange quark components for the first time. 

\vspace{-0.3cm}
\paragraph*{Gluon.}
The polarized gluon distribution has been elusive for a long time, since it is only 
weakly constrained by DIS and SIDIS measurements, where 
it enters as a higher-order correction, and by the limited $Q^2$ coverage of 
the data, in which sensitivity can come from the DGLAP evolution equations.
In fact, it was believed to be small until very recently. 
The availability of precise jet, di-jet, and 
hadron production spin asymmetry measurements at RHIC were a game changer
in this respect, revealing for the first time a sizable polarization of the gluon PDF
in the DSSV14 and in the NNPDFpol1.1 analyses. Such evidence,
however, is limited to the region $0.02\lesssim x\lesssim 0.4$, outside of which 
the polarized gluon PDF is affected by large extrapolation uncertainties that prevent
any definitive conclusion about its role in understanding the proton
spin decomposition (see Sect~\ref{susec:nonperturbative}
and Ref.~\cite{Stratmann:2020xxx}).

\vspace{-0.3cm}
\subsection{Nuclear PDFs}
\label{subsec:nuclear}

In the nuclear sector, the most recent PDF determinations are 
nCTEQ15~\cite{Kovarik:2015cma}, EPPS16~\cite{Eskola:2016oht},
nNNPDF1.0~\cite{AbdulKhalek:2019mzd} and TUJU19~\cite{Walt:2019slu}.
The nCTEQ15 analysis has since been updated with measurements
of vector boson production~\cite{Kusina:2016fxy} 
in proton-lead and lead-lead
collisions, while the EPPS collaboration recently
studied the impact of di-jet~\cite{Eskola:2019dui} and $D$-meson 
production~\cite{Eskola:2019bgf} in proton-lead collisions.

The differences among these PDF sets are summarized in Table~\ref{tab:pdfs_all}.
Concerning the data set, they are all based on inclusive DIS 
data except for neutrino DIS data in the case of nNNPDF1.0. 
Complementary measurements are 
added only in the nCTEQ15 and EPPS16 global analyses, where 
various fixed-target and collider DY observables are included
with jet and hadron production data. Finally, all of the PDF sets 
are determined at NLO, with the DIS-only 
nNNPDF1.0 and TUJU19 analyses extending also to NNLO.

%-------------------------------------------------------------------------------
\begin{figure}[!t]
\centering
\includegraphics[width=\textwidth]{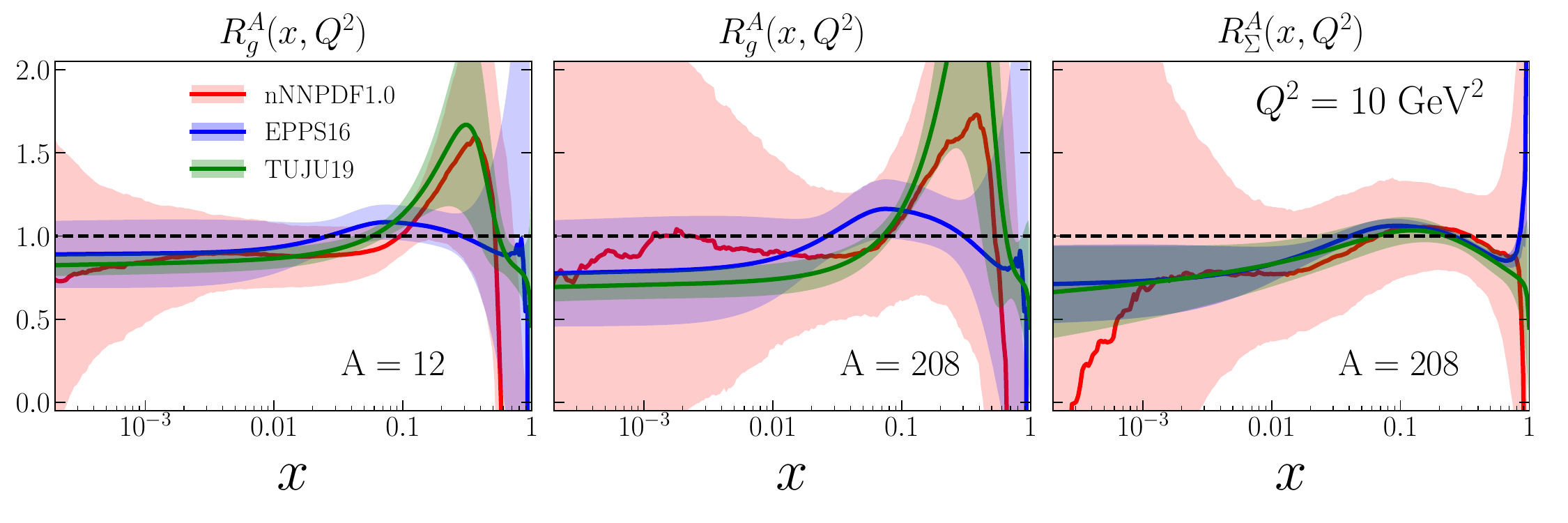}\\
\caption{Nuclear modification ratios (Eq.~\eqref{eq:Rmodification}) 
as a function of $x$ at $Q^2=10$~GeV$^2$
for the gluon PDF in $^{12}_{\ \, 6}{\rm C}$ and $^{208}_{\ \, 82}{\rm Pb}$,
and the quark singlet PDF in $^{208}_{\ \, 82}{\rm Pb}$. Results are shown 
for the nNNPDF1.0, EPPS16 and TUJU19 analyses at NLO. 
Uncertainty bands correspond to 90\% confidence levels.}
\label{fig:npdf_comparison}
\end{figure}
%-------------------------------------------------------------------------------

In the right panel of Fig.~\ref{fig:pdfs_all}, an example 
nuclear PDF set for lead is displayed at $Q^2=100$ GeV$^2$
as a function of $x$ for all flavors. 
Furthermore, we illustrate in Fig.~\ref{fig:npdf_comparison}
the ratio $R_f^A(x,Q^2)$~\eqref{eq:Rmodification} at $Q^2=10$ GeV$^2$ 
for the gluon distribution in a light nucleus, 
$^{12}_{\ 6}{\rm C}$, and a heavy nucleus, $^{208}_{\ 82}{\rm Pb}$. Also
shown is the 
quark singlet PDF combination for $^{208}_{\ 82}{\rm Pb}$. 
Here the nNNPDF1.0, EPPS16, and TUJU19 NLO PDF sets are 
compared with uncertainty bands corresponding to
90\% confidence levels.
The ratio $R_f^A(x,Q^2)$ presented in Fig.~\ref{fig:npdf_comparison}
provides the most relevant information in nuclear PDF studies
since it sheds light on the various nuclear modifications with respect to the
free proton in different $x$ regions. 
In particular, experimental data have 
indicated a shadowing behavior ($R<1$)
at low $x\lesssim0.1$, 
followed by anti-shadowing ($R>1$) in the $x\sim0.1$ region,
and lastly the EMC effect ($R<1$) in the valence region,
$x\sim0.4$, the latter being named after the experimental group from 
which it was discovered~\cite{Aubert:1983xm}. 
While the mechanisms that generate these
effects are still under investigation, they are expected
to appear as general features in the nuclear PDF results, which 
are summarized as follows.

\vspace{-0.3cm}
\paragraph*{Quarks.}
In Fig.~\ref{fig:pdfs_all}, 
the behavior of the valence and sea quark PDFs
mimic the distributions of their proton counterparts, albeit they are 
affected by larger uncertainties which tend to increase as the 
atomic number $A$ increases. The reason for this can be traced 
back to the quantity and quality of the data, which is significantly more 
limited in the variety of processes and in the breadth of the kinematic 
coverage for nuclear PDFs than for proton PDFs (refer to Fig.~\ref{fig:kinematics}). 
Note that, opposite to the proton case,
the up and down quark valence distributions are fairly similar
as a result of the nuclear PDF definition (Eq.~\eqref{eq:nucPDFs}).
Assuming exact isospin symmetry, the visible 
differences then arise from the 
non-isoscalarity of the nucleus. In general, the agreement between 
the different nuclear PDF sets 
is excellent for the total quark singlet distribution presented in 
Fig.~\ref{fig:npdf_comparison}, where the shape clearly 
displays shadowing, anti-shadowing, and EMC effects in their expected
$x$ regions. Additional comparisons of individual quark and antiquark 
flavors are not made here due to the lack of flavor separation in DIS 
data used in the nNNPDF1.0 and TUJU19 analyses, but 
can be found in Refs.~\cite{Eskola:2016oht,Kovarik:2015cma}. 
Lastly, despite containing similar datasets, the nNNPDF1.0 distributions show 
significantly larger uncertainties than the TUJU19 result, 
especially in the small-$x$ extrapolation region. While this is partly
due to the different uncertainty estimation techniques, it can also
be attributed to the use of a more flexible parameterization 
in nNNPDF1.0 (see Refs.~\cite{AbdulKhalek:2019mzd} 
and~\cite{Walt:2019slu} for more details).

\vspace{-0.3cm}
\paragraph*{Gluon.}
The features of the gluon distribution in Fig.~\ref{fig:pdfs_all} 
are also similar to those in the proton case, apart from having 
larger uncertainties. However,
the nuclear modifications to the gluon distribution 
in Fig.~\ref{fig:npdf_comparison}
are much less distinct than those for the quark singlet. 
While the TUJU19 PDF set displays very clear shadowing
behavior for the ratio, the larger uncertainties for the 
EPPS16 and nNNPDF1.0 distributions 
prevent any definitive conclusion about nuclear shadowing in
the low $x$ region. Furthermore, gluon anti-shadowing seems to
appear in different regions of $x$ for the different PDF analyses. 
For EPPS16, the gluon exhibits anti-shadowing
in a similar region as the quark singlet, while the 
TUJU19 and nNNPDF1.0 analyses display an anti-shadowing
effect at larger $x$. The latter is likely an artefact
of the fitting procedure rather than an actual physical effect, as the 
two analyses do not include data sensitive to the gluon. However,
further investigation is required to make any conclusive statements.  
Lastly, there is a noticeable difference in uncertainties between
the gluon uncertainties in carbon compared
to that of heavier lead nuclei. In fact, this is also true
for the quark distributions and arises by imposing the 
boundary condition at $A=1$ (see Sect.~\ref{subsec:evolution}).
In this sense, the proton PDFs can play a crucial role 
in constraining the nuclear distributions for light nuclei.

\vspace{-0.4cm}
\section{PRESENT AND FUTURE RELEVANCE OF PDFS}
\label{sec:relevance}

While significant progress has been achieved in the determination of PDFs, 
various issues remain open. In this section we give a concise summary
of the ones that we find more relevant in particle and hadronic physics,
and how future colliders will aid in solving them.

\subsection{Precision physics in the SM and beyond}
Since the discovery of the Higgs boson~\cite{Aad:2012tfa,*Chatrchyan:2012xdj},
to which an accurate determination of unpolarized PDFs was instrumental,
the LHC has allowed the SM to be established with unprecedented precision.
Nevertheless, PDFs remain one of the main sources of uncertainty in 
pushing forward precision and discovery physics at the LHC.

\vspace{-0.3cm}
\paragraph*{Determination of the strong coupling.}
It is customary to determine the value of the strong coupling $\alpha_s$ from
a variety of processes involving hadrons in the initial state, which require
knowledge of PDFs. In this respect, determinations of $\alpha_s$ have 
historically been of two kinds (see~\cite{Tanabashi:2018oca} for a review). 
The first realizes a simultaneous fit of both $\alpha_s$ and the PDFs in a 
global QCD analysis from a (more or less) wide set of data and 
processes; the second finds the likelihood of some new data for a single process
as a function of $\alpha_s$, based on a pool of pre-existing PDF sets determined
with various input values of the strong coupling. While it has been recently 
argued that only an appropriate implementation of the 
first method, which takes into account correlations between $\alpha_s$ and the 
PDFs, leads to an unbiased result~\cite{Forte:2020pyp}, in both cases 
the PDF uncertainty becomes relevant. For instance, at NNLO and at 
the $Z$-boson mass $m_Z$, the analysis of Ref.~\cite{Ball:2018iqk} found
$\alpha_s(m_Z)=0.1185\pm0.0005^{\rm exp}\pm0.0001^{\rm meth}\pm0.0011^{\rm th}$,
where exp, meth, and th uncertainties are the PDF, methodological and 
missing higher order theoretical uncertainties, respectively.

\vspace{-0.3cm}
\paragraph*{Determination of SM parameters.} 
Parton distributions represent one of the dominant sources of theoretical 
uncertainty for the determination of the Higgs boson couplings and 
cross sections at the LHC (see~\cite{Cepeda:2019klc} and references therein for 
a review). Likewise, they are the most significant limiting factor in the 
precise 
determination of electroweak parameters such as the mass of the $W$-boson, the 
mass of the top-quark, and the CKM matrix elements (see~\cite{Azzi:2019yne} and 
references therein, and~\cite{Bagnaschi:2019mzi} for a dedicated 
study).

\vspace{-0.3cm}
\paragraph*{The search for BSM physics.}
Beside precise benchmarks of the SM, extensive searches are in progress at the 
LHC for physics signatures beyond the SM. Since the majority of measurements
at the LHC and elsewhere have so far
been consistent with SM predictions, many have turned towards 
analyzing LHC data in the context of a SM effective field theory (SMEFT), 
where new physics signals may be found in the dynamics of SM fields via higher 
(mass) dimension operators (see Ref.~\cite{Brivio:2017vri} for a review). 
A large number of SMEFT analyses have been performed in both the electroweak 
and Higgs sectors, as well as the top quark sector, to constrain the Wilson 
coefficients related to the higher mass dimension terms. Here the PDFs play 
an essential role in the calculations of the SM part of the SMEFT expansion
(see {\it e.g.}~\cite{Hartland:2019bjb} and references therein). 
In principle, however, effects beyond the SM could be absorbed by the 
PDFs themselves in a global QCD analysis. To avoid this, one should
determine a set of SMEFT coefficients simultaneously with the PDFs.
Despite being a daunting task, it was demonstrated to 
be achievable in a simple case, whereby a small subset of SMEFT coefficients was 
determined together with the PDFs by analyzing inclusive DIS 
data~\cite{Carrazza:2019sec}. In that case, no significant indication of 
physics beyond the SM were found. However, much work still remains
in the field of SMEFT to obtain better constraints on the Wilson coefficients 
and identify new physics signals. As higher order QCD and electroweak 
calculations are made available in the SMEFT, and experimental measurements 
increase in abundance and precision, the role of PDFs in beyond the SM searches 
will become ever more relevant. 

\subsection{A window to non-perturbative behavior}
\label{susec:nonperturbative}

Because of their nature, PDFs can shed light on the non-perturbative aspects 
of nucleons and nuclei at perturbative scales. Among these, some of the most
compelling are as follows.

\vspace{-0.3cm}
\paragraph*{Sea quark asymmetries.} 
Perturbative QCD predicts light sea quarks to be equally produced by gluons in 
virtual fermion loops. However, a sizable asymmetry in the unpolarized anti-up and 
anti-down quark sea was observed in DY by the 
NA51~\cite{Amaudruz:1991at} experiment and confirmed by the 
NuSea/E866~\cite{Towell:2001nh} experiment at CERN and 
Fermilab, respectively. Such an asymmetry therefore has a non-perturbative origin, which 
is commonly explained in the context of a meson cloud 
model~\cite{Melnitchouk:1991ui,*Kumano:1991em,*Henley:1990kw,*Speth:1996pz}. 
Significant effort is underway to further constrain the $\bar{d}-\bar{u}$ 
unpolarized PDF asymmetry and also to clarify its origin, see {\it e.g.} 
Ref.~\cite{Ethier:2018efr} for a recent study of the meson cloud applied
to the $\Delta$ baryon. Experimentally, new DY measurements from the SeaQuest/E-906 
experiment at Fermilab~\cite{Aidala:2017ofy} are soon expected to be finalized.
Significant interest in flavor symmetry breaking effects exists also for other 
sea quark distributions, specifically the unpolarized (polarized) $s-\bar{s}$ 
($\Delta s-\Delta\bar{s}$) and polarized $\Delta \bar{u}-\Delta \bar{d}$ 
asymmetries. The former relates to constraining the strange quark 
PDF uncertainties, as discussed in Sect.~\ref{sec:results}.
The latter can be fairly well constrained by hadron multiplicities in SIDIS
and by $W^\pm$ single-spin asymmetries in polarized proton-proton 
collisions. Interestingly, these have shown evidence of a positive 
$\Delta\bar{u}-\Delta\bar{d}$ asymmetry in the intermediate $x$ region,
opposite to that generated in the unpolarized 
case~\cite{Nocera:2017wep,*Nocera:2014rea}. 

\vspace{-0.3cm}
\paragraph*{Down to up quark ratio at large x.}
Quarks carrying large momentum fractions can provide additional insight into 
non-perturbative quark-gluon interactions. Despite progress in PDF 
determinations, the $d/u$ and $\Delta d/\Delta u$ ratios are affected by large 
uncertainties that prevent any conclusive comparison with 
theoretical models~\cite{Ball:2016spl,Accardi:2011fa,*Owens:2012bv,
*Arrington:2011qt}. The reason for this state of affairs is the lack of 
experimental data with free neutron targets (or beams). Instead, light nuclei 
such as deuterium or helium are commonly used in (DIS) experiments, and therefore
bound nucleon effects must be taken into account in PDF analyses 
(see {\it e.g.} Ref.~\cite{Tropiano:2018quk} and references therein).
The CJ collaboration recently focused on the determination of the $u$ and $d$ 
PDFs at large $x$ in an analysis where nuclear effects in electron-nuclei 
scattering were treated by means of a weak binding approximation and off-shell 
corrections~\cite{Accardi:2016qay}. On the experimental
front, the 12 GeV program at JLab~\cite{Dudek:2012vr} is well underway with 
several dedicated 
experiments aiming to reveal large-$x$ PDF dynamics. 
In particular, the MARATHON 
experiment has recently completed measurements of unpolarized DIS from a 
tritium target, which aims to extract the neutron to proton structure function 
ratio up to $x\sim0.85$~\cite{MARATHON}. Results are expected 
to be reported soon, which can subsequently be used in a 
global QCD analysis to further reduce the $d/u$ ratio uncertainty. 

\vspace{-0.3cm}
\paragraph*{Non-perturbative charm.}
Apart from stabilizing predictions for a class of unpolarized hadronic cross 
sections at high energies, non-perturbative charm 
in the unpolarized proton is also relevant at low energies. Its role is 
primarily discussed in the context of a non-zero 5-quark Fock state component 
of the proton wave function, $|uudc\bar{c}\rangle$~\cite{Brodsky:2015fna}. 
Phenomenologically, world scattering data sensitive to the unpolarized charm 
quark PDF, most notably those measuring the $F_2^c$ structure function, can be 
analyzed by separating the charm component into a perturbative and a 
non-perturbative piece, the latter of which carries some fraction of the proton's 
momentum, 
$\langle x\rangle_{{\rm IC}}\equiv\int_0^1dx x\left[c(x)+\bar{c}(x)\right]\neq 0$.
Various phenomenological analyses suggest that this component is small
but sizable. The CTEQ-TEA collaboration found 
$\langle x \rangle_{\rm IC} \lesssim 2.5$\% at $Q=1.65$ GeV using various 
non-perturbative models~\cite{Dulat:2013hea}, while the NNPDF collaboration found
$\langle x \rangle_{\rm IC}=(1.6\pm 1.2)$\% at the same energy scale, in a fit 
where the charm PDF was parameterized together with the light quark 
PDFs~\cite{Ball:2016neh}.
On the other hand, the JR collaboration found $\langle x \rangle_{\rm IC} = (0.15\pm0.09)$\%, 
utilizing data at low final-state invariant-mass energies from 
SLAC~\cite{Jimenez-Delgado:2014zga}. While the NNPDF result is compatible with
the other two within its large uncertainties, it corresponds to a completely
different shape of the charm PDF.  Such differences are likely to 
originate from the various theoretical and 
methodological choices that are inherent in each analysis.

\vspace{-0.3cm}
\paragraph*{The proton spin.}
The size of the contribution of quarks, antiquarks and gluons to the nucleon 
spin is quantified by the first moments of the corresponding polarized PDFs,
according to the canonical decomposition of the proton's total angular 
momentum~\cite{Leader:2013jra}. Both the NNPDFpol1.1 and the DSSV14 
analyses agree on the values of the quark singlet and gluon polarized PDF 
first moments in the data region, which are found to be 0.23-0.30 and 0.20-0.23 
at $Q^2=10$ GeV$^2$, respectively. 
However, uncertainties affecting the small-$x$ extrapolation region,
as well as those surrounding SU$_f$ symmetry constraints, 
prevent an assessment of the residual
contribution to the total angular momentum of the proton coming from
the orbital angular momentum of quarks and gluons.

\vspace{-0.3cm}
\subsection{Parton dynamics in the nuclear medium}
The relevance of nuclear parton distributions is perhaps the most far-reaching 
of the collinear distributions, especially in the following domains.

\vspace{-0.3cm}
\paragraph*{Nuclear modifications and the characterization of quark-gluon plasma.}
Nuclear PDFs are key to understanding how parton dynamics are modified 
in the nuclear medium and the mechanisms that generate the
shadowing, anti-shadowing, and EMC effects~\cite{Hen:2013oha}.
In addition, they play an essential role in the 
characterization of the quark-gluon plasma, the hot and dense medium created in 
the early Universe currently being studied by heavy ion collisions at RHIC 
and LHC~\cite{Abreu:2007kv,*Adams:2005dq}. In this respect, measurements
of proton-ion and ion-ion collisions will help constrain the nuclear PDFs and
their uncertainties. In addition to those outlined in Sect.~\ref{sec:expdata}, 
the LHCb $D$-meson production data were demonstrated to constrain the nuclear 
gluon PDF at low $x$~\cite{Eskola:2019bgf}. More recently, measurements of 
$J/\psi$ production in ultra-peripheral collisions of two lead nuclei were made 
available 
by the ALICE experiment at the LHC~\cite{Acharya:2019vlb}; similar measurements 
of gold nuclei are expected to be made by the STAR experiment at 
RHIC~\cite{Adam:2019rxb}. Such collisions, where the impact parameter is larger 
than the combined radii of the colliding nuclei, provide a rather clean way to 
study photon-nucleus interactions. While coherent photoproduction of vector 
mesons can provide auxillary constraints on nuclear PDFs in general,
$J/\psi$ is of special interest since it is a gluon initiated process 
and therefore its associated theoretical prediction is sensitive to the nuclear gluon 
PDF~\cite{Ryskin:1992ui}. Once included in a PDF determination, the data can 
potentially provide additional evidence of shadowing and gluon saturation at small $x$ values. 

\vspace{-0.3cm}
\paragraph*{Relationship with proton PDFs.}
A significant number of observables used in global QCD analyses of 
free-proton PDFs involve proton-nucleus or lepton-nucleus scattering, 
namely proton-deuteron (and proton-ion) DY measurements,
and charged-current neutrino-nucleus measurements. Being essential to 
free-proton PDF extractions, in particular to constrain the sea quark PDFs, 
nuclear corrections must be carefully taken into 
account using a nuclear model or parameterized and determined from a
fit to the data. Whatever approach is adopted, 
effects from nuclei will consequently
increase the uncertainty in proton PDFs since 
nuclear corrections are known less precisely. A methodology to include such an
uncertainty in proton fits was developed in~\cite{Ball:2018twp}, 
as was discussed in Sect.~\ref{subsec:uncertainties}.

\vspace{-0.3cm}
\paragraph*{Implications for astroparticle physics.}
Neutrino telescopes, such as IceCube~\cite{IceCube:2018cha},
KM3NeT~\cite{Adrian-Martinez:2016fdl} and Baikal-GVD~\cite{Avrorin:2018ijk}, 
have been developed in recent years to measure ultra high energy neutrino
fluxes (with energy $E_\nu\gtrsim 10^7$~GeV) as a way to study, in conjunction 
with cosmic ray observatories such as Auger~\cite{Aab:2019gra}, 
sources of cosmic rays in the Universe and QCD at multi-TeV scales, far
outside the energy scales probed by available colliders.
Theoretical predictions of neutrino-nucleus scattering cross 
sections can be made in the framework of perturbative QCD for both 
signal~\cite{Bertone:2018dse} and backround~\cite{Bhattacharya:2016jce} events. 
Since the detectors measure scattering events from a target source material 
consisting of nuclei, usually water molecules, a precise knowledge of nuclear 
PDFs is necessary. Nuclear modifications induced by the medium increase the 
cross sections by 1-2\% at energies below 100~TeV (antishadowing), and reduce 
it by 3-4\% at higher energies (shadowing)~\cite{Klein:2020nuk}. Nuclear 
effects also alter the inelasticity distribution of neutrino interactions in 
water/ice by increasing the number of low inelasticity interactions, with a 
larger effect for neutrinos than antineutrinos. These effects are particularly
important in the energy range below a few TeV. An accurate control of nuclear
PDF uncertainties is pivotal to characterize all of these effects, in 
particular above the IceCube energy reach, still within the larger reach
of Auger ($E_\nu\sim 10^{12}$~GeV), where the absence of experiment constraints
remains the limiting factor to determine the composition of the highest energy 
hadronic cosmic rays~\cite{Henley:2005ms}. 

\subsection{The role of future colliders}
\label{subsec:future}

Several new accelerator upgrades and designs are being planned that will 
improve much of the PDF statuses discussed in the above sections. The LHC is 
currently upgrading to its high luminosity (HL) 
phase~\cite{Apollinari:2017cqg}, where various observables will be measured 
with significantly increased statistical precision. This will extend the LHC 
physics output for another decade, providing rich precision measurements of SM 
parameters such as the Higgs couplings and the $W$-boson 
mass~\cite{Azzi:2019yne,Cepeda:2019klc}. At lower
energies, the JLab has just completed its upgrade to 12 GeV~\cite{Dudek:2012vr},
which will allow for a careful investigation of aspects related to the non-perturbative 
structure of the unpolarized and polarized proton.

Beside these upgrades, there are at least two additional accelerator
designs being developed that will be particularly beneficial for PDF studies. 
The first is an electron-proton collider at the LHC (LHeC), which intends
to inject roughly 60 GeV electron beams into the LHC proton 
collider~\cite{AbelleiraFernandez:2012cc}. With the point-like nature of the 
electron probe and the energies of the LHC, high precision DIS measurements 
will be achieved for Bjorken-$x$ down to $\sim5\times10^{-6}$,
a kinematic region that has yet to be explored.
The second is an electron-ion collider (EIC) planned to be built on the 
current site of RHIC, with medium energies of up to 
90 GeV~\cite{Accardi:2012qut,*Boer:2011fh}. The aim of the EIC is to study with 
high precision the three-dimensional structure of nucleons and nuclei. In 
terms of collinear structure, however, the new accelerator will be able to 
provide high precision DIS measurements for nuclear PDF determination down to 
$x\sim10^{-4}$. One of the physics aims, in fact, is to provide such
measurements in order to pin down the nuclear gluon distribution and reveal
gluon saturation effects, similarly to what HERA did for the proton.
It is also a machine that can polarize the beams, 
thus providing important constraints for the polarized PDFs.

The extended kinematic regions attained by both the LHeC and the EIC are
displayed as dashed regions in Fig.~\ref{fig:kinematics}. Impact studies
of projected pseudodata foreseen at these facilities have been extensively 
performed for unpolarized PDFs in 
Ref.~\cite{AbdulKhalek:2019mps,*Khalek:2018mdn,Aschenauer:2019kzf},
for polarized PDFs in Refs.~\cite{Aschenauer:2012ve,*Aschenauer:2015ata,
*Aschenauer:2013iia,Ball:2013tyh}, and for nuclear PDFs in 
Refs.~\cite{HannuPaukkunenfortheLHeCstudygroup:2017ric,*Helenius:2015xda,
Aschenauer:2017oxs,AbdulKhalek:2019mzd}.

\vspace{-0.4cm}
\section{OUTLOOK}
\label{sec:outlook}

Global QCD analyses of PDFs continue to be an active 
field in particle and hadronic physics. The upgrade of the LHC,
combined with the advent of new colliders,
will enhance their relevance even further in the next decades. 
Now that PDFs are entering a new precision era, they will
be critical for calculations of signal and background events in 
physics searches beyond the SM and
in bringing electroweak symmetry breaking under complete
control. They will allow us to explore nuclear modifications, the 
onset of gluon saturation, and aid in understanding 
high-energy neutrino interactions in astroparticle physics. 
And they will shed light on the 
non-perturbative structure of the proton and 
unravel the proton's spin decomposition.
To achieve such ambitious goals, we foresee more sophisticated 
PDF determinations in the future, all of which 
should fulfill the following criteria.

\begin{summary}[]
\begin{enumerate}

\item PDF determinations should be as global as possible, {\it i.e.} the range
of the input dataset should be extended to cover unexplored kinematic 
regions and new processes. In this respect, experimental uncertainties
should be carefully scrutinized, in particular as correlated
systematic uncertainties dominate collider measurements. The observed tensions
between data sets, and the different parton sensitivity to different observables 
across PDF extractions must be carefully understood with appropriate statistical tools.
Benchmarking (and, if beneficial, combination) of different PDF sets, including 
polarized and nuclear PDFs, must become standard.

\item Global QCD analyses should be as accurate as possible, {\it i.e.} 
theoretical calculations should be performed at the highest available order
in perturbative QCD. This is currently NNLO for unpolarized and nuclear PDFs,
and NLO for polarized PDFs. Approximate higher-order PDF fits can actually be 
performed only to inclusive DIS data. In this respect, a N$^3$LO determination 
of unpolarized PDFs is desirable to study the effect of using a NNLO matching 
condition, and whether it leads to a charm PDF closer to the fitted one.
A NNLO determination of polarized PDFs, instead, will allow us to check the 
perturbative stability of all of the current polarized PDF sets. 
Unpolarized PDF determinations should also resum large logarithms:
large-$x$ resummation is particularly important for searches of new 
physics and matching of fixed-order calculations to parton showering in Monte 
Carlo event generators, while small-$x$ resummation is
relevant to reveal the onset of BFKL dynamics and gluon saturation.
Finally, the treatment of heavy quark PDFs on the same footing as 
the light quark distributions, and also the systematic 
inclusion of electroweak corrections, should 
become standard for unpolarized PDFs.

\item PDF determinations should represent uncertainties as faithfully as 
possible, {\it i.e.} their statistical meaning should be carefully validated. 
Furthermore, a complete characterization of 
theoretical uncertainties should be implemented. This includes 
the uncertainty from input physical 
parameters and from the neglect of missing higher orders or
other corrections in the calculation of hadronic observables (such as 
nuclear and higher-twist corrections). In view of related needs for 
precision physics, this will become mandatory for unpolarized PDFs.

\item PDF analyses should aim at becoming simultaneous, {\it i.e.}
parameters from different non-perturbative aspects of the theory should be 
determined at the same time. The simplest example is represented by the 
determination of the strong coupling along with the PDFs. More ambitiously, we 
envision three cases where simultaneous fits will become critical. The
first is in the extraction of polarized PDFs, where observables are 
often presented as ratios of spin-dependent to spin-averaged cross sections.
In this case, unpolarized and polarized PDFs should be treated
simultaneously in a universal QCD analysis.
The second is the analysis of semi-inclusive measurements involving the 
production of an identified hadron in the final state. We believe that these 
processes can be analyzed accurately only if the final-state FFs are determined 
together with the initial-state PDFs. While this is primarily significant for QCD analyses of 
polarized PDFs, it can also provide additional information on 
the sea distributions in unpolarized PDF fits.
The third is analyses of LHC processes in the SMEFT. We believe 
that searches for physics beyond the SM can be realized in this
framework only if SMEFT Wilson coefficients are extracted simultaneously with
the PDF parameters. Because such fits will dramatically 
increase the complexity of 
the parameter space, development of 
new optimization techniques such as those based on 
machine learning and artificial intelligence will likely be required.

\item PDF extractions should benefit from input provided by 
{\it ab-initio} lattice QCD calculations~\cite{Lin:2017snn}
and by advancements in the theoretical and phenomenological understanding 
of non-perturbative functions that generalize collinear 
PDFs~\cite{Angeles-Martinez:2015sea,*Bacchetta:2016ccz}. These fields
witnessed spectacular progress in recent years, which cannot be ignored.

\end{enumerate}
\end{summary}

As can be seen by this review, significant efforts are underway to 
achieve many of the objectives listed here. While much work 
remains, global QCD analyses have undoubtably taken major steps 
forward to address challenging questions in QCD and deepen 
our understanding of nucleon and nuclei substructure.

\vspace{-0.4cm}
%Disclosure
\section*{DISCLOSURE STATEMENT} 
The authors are not aware of any affiliations, memberships, funding, 
or financial holdings that might be perceived as affecting the objectivity 
of this review. 

\vspace{-0.4cm}
%Funding
\section*{ACKNOWLEDGMENTS}
We are indebted to all members of the NNPDF collaboration, in particular 
R.D.~Ball, S.~Forte and J.~Rojo. J.E. is supported by the Netherlands 
Organization for Scientific Research (NWO). E.R.N. is funded by the European 
Commission through the Marie Sk\l odowska-Curie Action ParDHonS\_FFs.TMDs 
(grant number 752748). 

\vspace{-0.4cm}
\bibliographystyle{ar-style5}
\bibliography{Ethier_Nocera_review}

\begin{mcbibliography}{240}
\expandafter\ifx\csname natexlab\endcsname\relax\def\natexlab#1{#1}\fi

\bibitem{Gao:2017yyd}
Gao J, Harland-Lang L, Rojo J.
\newblock \textit{Phys. Rept.} 742:1 (2018)\relax
\relax
\bibitem{Kovarik:2019xvh}
Kovařík K, Nadolsky PM, Soper DE  arXiv:1905.06957 [hep-ph] (2019)\relax
\relax
\bibitem{Aidala:2012mv}
Aidala CA, Bass SD, Hasch D, Mallot GK.
\newblock \textit{Rev. Mod. Phys.} 85:655 (2013)\relax
\relax
\bibitem{Forte:2013wc}
Forte S, Watt G.
\newblock \textit{Ann. Rev. Nucl. Part. Sci.} 63:291 (2013)\relax
\relax
\bibitem{Jimenez-Delgado:2013sma}
Jimenez-Delgado P, Melnitchouk W, Owens JF.
\newblock \textit{J. Phys.} G40:093102 (2013)\relax
\relax
\bibitem{Collins:1989gx}
Collins JC, Soper DE, Sterman GF.
\newblock \textit{Adv. Ser. Direct. High Energy Phys.} 5:1 (1989)\relax
\relax
\bibitem{Gross:1973id}
Gross DJ, Wilczek F.
\newblock \textit{Phys. Rev. Lett.} 30:1343 (1973)\relax
\relax
\bibitem{Politzer:1973fx}
Politzer HD.
\newblock \textit{Phys. Rev. Lett.} 30:1346 (1973)\relax
\relax
\bibitem{Ellis:1991qj}
Ellis RK, Stirling WJ, Webber BR.
\newblock vol.~8 (1996)\relax
\relax
\bibitem{Collins:2011zzd}
Collins J.
\newblock vol.~32 (2011)\relax
\relax
\bibitem{Blumlein:2012bf}
Blumlein J.
\newblock \textit{Prog. Part. Nucl. Phys.} 69:28 (2013)\relax
\relax
\bibitem{deFlorian:1997zj}
de~Florian D, Stratmann M, Vogelsang W.
\newblock \textit{Phys. Rev.} D57:5811 (1998)\relax
\relax
\bibitem{Metz:2016swz}
Metz A, Vossen A.
\newblock \textit{Prog. Part. Nucl. Phys.} 91:136 (2016)\relax
\relax
\bibitem{Barshay:1975zz}
Barshay S, Dover CB, Vary JP.
\newblock \textit{Phys. Rev.} C11:360 (1975)\relax
\relax
\bibitem{tHooft:1972tcz}
't~Hooft G, Veltman MJG.
\newblock \textit{Nucl. Phys.} B44:189 (1972)\relax
\relax
\bibitem{tHooft:1973mfk}
't~Hooft G.
\newblock \textit{Nucl. Phys.} B61:455 (1973)\relax
\relax
\bibitem{Bardeen:1978yd}
Bardeen WA, Buras AJ, Duke DW, Muta T.
\newblock \textit{Phys. Rev.} D18:3998 (1978)\relax
\relax
\bibitem{Herzog:2017ohr}
Herzog F, et~al.
\newblock \textit{JHEP} 02:090 (2017)\relax
\relax
\bibitem{Luthe:2017ttg}
Luthe T, Maier A, Marquard P, Schroder Y.
\newblock \textit{JHEP} 10:166 (2017)\relax
\relax
\bibitem{Herzog:2018kwj}
Herzog F, et~al.
\newblock \textit{Phys. Lett.} B790:436 (2019)\relax
\relax
\bibitem{Gribov:1972ri}
Gribov VN, Lipatov LN.
\newblock \textit{Sov. J. Nucl. Phys.} 15:438 (1972), [Yad.
  Fiz.15,781(1972)]\relax
\relax
\bibitem{Lipatov:1974qm}
Lipatov LN.
\newblock \textit{Sov. J. Nucl. Phys.} 20:94 (1975), [Yad.
  Fiz.20,181(1974)]\relax
\relax
\bibitem{Altarelli:1977zs}
Altarelli G, Parisi G.
\newblock \textit{Nucl. Phys.} B126:298 (1977)\relax
\relax
\bibitem{Dokshitzer:1977sg}
Dokshitzer YL.
\newblock \textit{Sov. Phys. JETP} 46:641 (1977), [Zh. Eksp. Teor.
  Fiz.73,1216(1977)]\relax
\relax
\bibitem{Moch:2004pa}
Moch S, Vermaseren JAM, Vogt A.
\newblock \textit{Nucl. Phys.} B688:101 (2004)\relax
\relax
\bibitem{Vogt:2004mw}
Vogt A, Moch S, Vermaseren JAM.
\newblock \textit{Nucl. Phys.} B691:129 (2004)\relax
\relax
\bibitem{Ablinger:2017tan}
Ablinger J, et~al.
\newblock \textit{Nucl. Phys.} B922:1 (2017)\relax
\relax
\bibitem{Moch:2017uml}
Moch S, et~al.
\newblock \textit{JHEP} 10:041 (2017)\relax
\relax
\bibitem{Moch:2015usa}
Moch S, Vermaseren JAM, Vogt A.
\newblock \textit{Phys. Lett.} B748:432 (2015)\relax
\relax
\bibitem{Vogt:2008yw}
Vogt A, Moch S, Rogal M, Vermaseren JAM.
\newblock \textit{Nucl. Phys. Proc. Suppl.} 183:155 (2008)\relax
\relax
\bibitem{Mitov:2006ic}
Mitov A, Moch S, Vogt A.
\newblock \textit{Phys. Lett.} B638:61 (2006)\relax
\relax
\bibitem{Moch:2007tx}
Moch S, Vogt A.
\newblock \textit{Phys. Lett.} B659:290 (2008)\relax
\relax
\bibitem{Almasy:2011eq}
Almasy AA, Moch S, Vogt A.
\newblock \textit{Nucl. Phys.} B854:133 (2012)\relax
\relax
\bibitem{Ciafaloni:2003ek}
Ciafaloni M, et~al.
\newblock \textit{Phys. Lett.} B576:143 (2003)\relax
\relax
\bibitem{Ciafaloni:2003rd}
Ciafaloni M, Colferai D, Salam GP, Stasto AM.
\newblock \textit{Phys. Rev.} D68:114003 (2003)\relax
\relax
\bibitem{Fadin:1975cb}
Fadin VS, Kuraev EA, Lipatov LN.
\newblock \textit{Phys. Lett.} 60B:50 (1975)\relax
\relax
\bibitem{Kuraev:1976ge}
Kuraev EA, Lipatov LN, Fadin VS.
\newblock \textit{Sov. Phys. JETP} 44:443 (1976), [Zh. Eksp. Teor.
  Fiz.71,840(1976)]\relax
\relax
\bibitem{Kuraev:1977fs}
Kuraev EA, Lipatov LN, Fadin VS.
\newblock \textit{Sov. Phys. JETP} 45:199 (1977), [Zh. Eksp. Teor.
  Fiz.72,377(1977)]\relax
\relax
\bibitem{Balitsky:1978ic}
Balitsky II, Lipatov LN.
\newblock \textit{Sov. J. Nucl. Phys.} 28:822 (1978), [Yad.
  Fiz.28,1597(1978)]\relax
\relax
\bibitem{Catani:2014uta}
Catani S, et~al.
\newblock \textit{Nucl. Phys.} B888:75 (2014)\relax
\relax
\bibitem{Appelquist:1974tg}
Appelquist T, Carazzone J.
\newblock \textit{Phys. Rev.} D11:2856 (1975)\relax
\relax
\bibitem{Collins:1978wz}
Collins JC, Wilczek F, Zee A.
\newblock \textit{Phys. Rev.} D18:242 (1978)\relax
\relax
\bibitem{Laenen:1992zk}
Laenen E, Riemersma S, Smith J, van Neerven WL.
\newblock \textit{Nucl. Phys.} B392:162 (1993)\relax
\relax
\bibitem{Laenen:1992xs}
Laenen E, Riemersma S, Smith J, van Neerven WL.
\newblock \textit{Nucl. Phys.} B392:229 (1993)\relax
\relax
\bibitem{Berger:2016inr}
Berger EL, et~al.
\newblock \textit{Phys. Rev. Lett.} 116:212002 (2016)\relax
\relax
\bibitem{Gao:2017kkx}
Gao J.
\newblock \textit{JHEP} 02:026 (2018)\relax
\relax
\bibitem{Ablinger:2017err}
Ablinger J, et~al.
\newblock \textit{Nucl. Phys.} B921:585 (2017)\relax
\relax
\bibitem{Ablinger:2019gpu}
Ablinger J, et~al.
\newblock \textit{Nucl. Phys.} B952:114916 (2020)\relax
\relax
\bibitem{Aivazis:1993pi}
Aivazis MAG, Collins JC, Olness FI, Tung WK.
\newblock \textit{Phys. Rev.} D50:3102 (1994)\relax
\relax
\bibitem{Collins:1998rz}
Collins JC.
\newblock \textit{Phys. Rev.} D58:094002 (1998)\relax
\relax
\bibitem{Guzzi:2011ew}
Guzzi M, Nadolsky PM, Lai HL, Yuan CP.
\newblock \textit{Phys. Rev.} D86:053005 (2012)\relax
\relax
\bibitem{Thorne:1997ga}
Thorne RS, Roberts RG.
\newblock \textit{Phys. Rev.} D57:6871 (1998)\relax
\relax
\bibitem{Thorne:1997uu}
Thorne RS, Roberts RG.
\newblock \textit{Phys. Lett.} B421:303 (1998)\relax
\relax
\bibitem{Thorne:2006qt}
Thorne RS.
\newblock \textit{Phys. Rev.} D73:054019 (2006)\relax
\relax
\bibitem{Cacciari:1998it}
Cacciari M, Greco M, Nason P.
\newblock \textit{JHEP} 05:007 (1998)\relax
\relax
\bibitem{Forte:2010ta}
Forte S, Laenen E, Nason P, Rojo J.
\newblock \textit{Nucl. Phys.} B834:116 (2010)\relax
\relax
\bibitem{Binoth:2010nha}
Binoth T, et~al. 2010.
\newblock In \textit{{Physics at TeV colliders. Proceedings, 6th Workshop,
  dedicated to Thomas Binoth, Les Houches, France, June 8-26, 2009}}\relax
\relax
\bibitem{Ball:2015tna}
Ball RD, et~al.
\newblock \textit{Phys. Lett.} B754:49 (2016)\relax
\relax
\bibitem{Ball:2015dpa}
Ball RD, Bonvini M, Rottoli L.
\newblock \textit{JHEP} 11:122 (2015)\relax
\relax
\bibitem{Forte:2019hjc}
Forte S, Giani T, Napoletano D.
\newblock \textit{Eur. Phys. J.} C79:609 (2019)\relax
\relax
\bibitem{Epele:2018ewr}
Epele M, García~Canal C, Sassot R.
\newblock \textit{Phys. Lett.} B790:102 (2019)\relax
\relax
\bibitem{Bertone:2015lqa}
Bertone V, Carrazza S, Pagani D, Zaro M.
\newblock \textit{JHEP} 11:194 (2015)\relax
\relax
\bibitem{Manohar:2016nzj}
Manohar A, Nason P, Salam GP, Zanderighi G.
\newblock \textit{Phys. Rev. Lett.} 117:242002 (2016)\relax
\relax
\bibitem{Manohar:2017eqh}
Manohar AV, Nason P, Salam GP, Zanderighi G.
\newblock \textit{JHEP} 12:046 (2017)\relax
\relax
\bibitem{Altarelli:1998gn}
Altarelli G, Forte S, Ridolfi G.
\newblock \textit{Nucl. Phys.} B534:277 (1998)\relax
\relax
\bibitem{Bjorken:1966jh}
Bjorken JD.
\newblock \textit{Phys. Rev.} 148:1467 (1966)\relax
\relax
\bibitem{Bjorken:1969mm}
Bjorken JD.
\newblock \textit{Phys. Rev.} D1:1376 (1970)\relax
\relax
\bibitem{Ethier:2017zbq}
Ethier JJ, Sato N, Melnitchouk W.
\newblock \textit{Phys. Rev. Lett.} 119:132001 (2017)\relax
\relax
\bibitem{Jaffe:1989jz}
Jaffe RL, Manohar A.
\newblock \textit{Nucl. Phys.} B337:509 (1990)\relax
\relax
\bibitem{Bass:2009ed}
Bass SD, Thomas AW.
\newblock \textit{Phys. Lett.} B684:216 (2010)\relax
\relax
\bibitem{AbdulKhalek:2019mzd}
Abdul~Khalek R, Ethier JJ, Rojo J.
\newblock \textit{Eur. Phys. J.} C79:471 (2019)\relax
\relax
\bibitem{Eskola:2016oht}
Eskola KJ, Paakkinen P, Paukkunen H, Salgado CA.
\newblock \textit{Eur. Phys. J.} C77:163 (2017)\relax
\relax
\bibitem{Giele:2001mr}
Giele WT, Keller SA, Kosower DA  arXiv:hep-ph/0104052 [hep-ph] (2001)\relax
\relax
\bibitem{Regge:1959mz}
Regge T.
\newblock \textit{Nuovo Cim.} 14:951 (1959)\relax
\relax
\bibitem{Brodsky:1973kr}
Brodsky SJ, Farrar GR.
\newblock \textit{Phys. Rev. Lett.} 31:1153 (1973)\relax
\relax
\bibitem{Ball:2016spl}
Ball RD, Nocera ER, Rojo J.
\newblock \textit{Eur. Phys. J.} C76:383 (2016)\relax
\relax
\bibitem{Nocera:2014uea}
Nocera ER.
\newblock \textit{Phys. Lett.} B742:117 (2015)\relax
\relax
\bibitem{Harland-Lang:2014zoa}
Harland-Lang LA, Martin AD, Motylinski P, Thorne RS.
\newblock \textit{Eur. Phys. J.} C75:204 (2015)\relax
\relax
\bibitem{Dulat:2015mca}
Dulat S, et~al.
\newblock \textit{Phys. Rev.} D93:033006 (2016)\relax
\relax
\bibitem{Glazov:2010bw}
Glazov A, Moch S, Radescu V.
\newblock \textit{Phys. Lett.} B695:238 (2011)\relax
\relax
\bibitem{Ball:2017nwa}
Ball RD, et~al.
\newblock \textit{Eur. Phys. J.} C77:663 (2017)\relax
\relax
\bibitem{Ball:2013lla}
Ball RD, et~al.
\newblock \textit{Nucl. Phys.} B874:36 (2013)\relax
\relax
\bibitem{Kovarik:2015cma}
Kovarik K, et~al.
\newblock \textit{Phys. Rev.} D93:085037 (2016)\relax
\relax
\bibitem{Alekhin:2017kpj}
Alekhin S, Blümlein J, Moch S, Placakyte R.
\newblock \textit{Phys. Rev.} D96:014011 (2017)\relax
\relax
\bibitem{Ball:2016neh}
Ball RD, et~al.
\newblock \textit{Eur. Phys. J.} C76:647 (2016)\relax
\relax
\bibitem{Bertone:2017bme}
Bertone V, Carrazza S, Hartland NP, Rojo J.
\newblock \textit{SciPost Phys.} 5:008 (2018)\relax
\relax
\bibitem{DAgostini:1993arp}
D'Agostini G.
\newblock \textit{Nucl. Instrum. Meth.} A346:306 (1994)\relax
\relax
\bibitem{Ball:2009qv}
Ball RD, et~al.
\newblock \textit{JHEP} 05:075 (2010)\relax
\relax
\bibitem{Vogt:2004ns}
Vogt A.
\newblock \textit{Comput. Phys. Commun.} 170:65 (2005)\relax
\relax
\bibitem{Salam:2008qg}
Salam GP, Rojo J.
\newblock \textit{Comput. Phys. Commun.} 180:120 (2009)\relax
\relax
\bibitem{Botje:2010ay}
Botje M.
\newblock \textit{Comput. Phys. Commun.} 182:490 (2011)\relax
\relax
\bibitem{Bertone:2013vaa}
Bertone V, Carrazza S, Rojo J.
\newblock \textit{Comput. Phys. Commun.} 185:1647 (2014)\relax
\relax
\bibitem{Bertone:2015cwa}
Bertone V, Carrazza S, Nocera ER.
\newblock \textit{JHEP} 03:046 (2015)\relax
\relax
\bibitem{Giele:2002hx}
Giele W, et~al. 2002.
\newblock In \textit{{Physics at TeV colliders. Proceedings, Euro Summer
  School, Les Houches, France, May 21-June 1, 2001}}\relax
\relax
\bibitem{Dittmar:2005ed}
Dittmar M, et~al.  arXiv:hep-ph/0511119 [hep-ph] (2005)\relax
\relax
\bibitem{Stratmann:2001pb}
Stratmann M, Vogelsang W.
\newblock \textit{Phys. Rev.} D64:114007 (2001)\relax
\relax
\bibitem{Bertone:2016lga}
Bertone V, Carrazza S, Hartland NP.
\newblock \textit{Comput. Phys. Commun.} 212:205 (2017)\relax
\relax
\bibitem{Carli:2010rw}
Carli T, et~al.
\newblock \textit{Eur. Phys. J.} C66:503 (2010)\relax
\relax
\bibitem{Wobisch:2011ij}
Wobisch M, et~al.  arXiv:1109.1310 [hep-ph] (2011)\relax
\relax
\bibitem{Campbell:2019dru}
Campbell J, Neumann T.
\newblock \textit{JHEP} 12:034 (2019)\relax
\relax
\bibitem{Nagy:2003tz}
Nagy Z.
\newblock \textit{Phys. Rev.} D68:094002 (2003)\relax
\relax
\bibitem{Alwall:2014hca}
Alwall J, et~al.
\newblock \textit{JHEP} 07:079 (2014)\relax
\relax
\bibitem{Bertone:2014zva}
Bertone V, et~al.
\newblock \textit{JHEP} 08:166 (2014)\relax
\relax
\bibitem{DelDebbio:2013kxa}
Del~Debbio L, Hartland NP, Schumann S.
\newblock \textit{Comput. Phys. Commun.} 185:2115 (2014)\relax
\relax
\bibitem{Ball:2014uwa}
Ball RD, et~al.
\newblock \textit{JHEP} 04:040 (2015)\relax
\relax
\bibitem{Bertone:2017tyb}
Bertone V, et~al.
\newblock \textit{Eur. Phys. J.} C77:516 (2017)\relax
\relax
\bibitem{Carrazza:2019mzf}
Carrazza S, Cruz-Martinez J.
\newblock \textit{Eur. Phys. J.} C79:676 (2019)\relax
\relax
\bibitem{Pumplin:2001ct}
Pumplin J, et~al.
\newblock \textit{Phys. Rev.} D65:014013 (2001)\relax
\relax
\bibitem{Stump:2001gu}
Stump D, et~al.
\newblock \textit{Phys. Rev.} D65:014012 (2001)\relax
\relax
\bibitem{Martin:2009iq}
Martin AD, Stirling WJ, Thorne RS, Watt G.
\newblock \textit{Eur. Phys. J.} C63:189 (2009)\relax
\relax
\bibitem{Ball:2010gb}
Ball RD, et~al.
\newblock \textit{Nucl. Phys.} B849:112 (2011), [Erratum: Nucl.
  Phys.B854,926(2012); Erratum: Nucl. Phys.B855,927(2012)]\relax
\relax
\bibitem{Ball:2011gg}
Ball RD, et~al.
\newblock \textit{Nucl. Phys.} B855:608 (2012)\relax
\relax
\bibitem{Hou:2016sho}
Hou TJ, et~al.
\newblock \textit{JHEP} 03:099 (2017)\relax
\relax
\bibitem{Carrazza:2015aoa}
Carrazza S, et~al.
\newblock \textit{Eur. Phys. J.} C75:369 (2015)\relax
\relax
\bibitem{Carrazza:2015hva}
Carrazza S, Latorre JI, Rojo J, Watt G.
\newblock \textit{Eur. Phys. J.} C75:474 (2015)\relax
\relax
\bibitem{Paukkunen:2014zia}
Paukkunen H, Zurita P.
\newblock \textit{JHEP} 12:100 (2014)\relax
\relax
\bibitem{Butterworth:2015oua}
Butterworth J, et~al.
\newblock \textit{J. Phys.} G43:023001 (2016)\relax
\relax
\bibitem{Schmidt:2018hvu}
Schmidt C, Pumplin J, Yuan CP, Yuan P.
\newblock \textit{Phys. Rev.} D98:094005 (2018)\relax
\relax
\bibitem{Carrazza:2016htc}
Carrazza S, Forte S, Kassabov Z, Rojo J.
\newblock \textit{Eur. Phys. J.} C76:205 (2016)\relax
\relax
\bibitem{Wang:2018heo}
Wang BT, et~al.
\newblock \textit{Phys. Rev.} D98:094030 (2018)\relax
\relax
\bibitem{Lin:2017stx}
Lin HW, et~al.
\newblock \textit{Phys. Rev. Lett.} 120:152502 (2018)\relax
\relax
\bibitem{Gbedo:2017eyp}
Gbedo YG, Mangin-Brinet M.
\newblock \textit{Phys. Rev.} D96:014015 (2017)\relax
\relax
\bibitem{Martin:2003sk}
Martin AD, Roberts RG, Stirling WJ, Thorne RS.
\newblock \textit{Eur. Phys. J.} C35:325 (2004)\relax
\relax
\bibitem{Accardi:2009br}
Accardi A, et~al.
\newblock \textit{Phys. Rev.} D81:034016 (2010)\relax
\relax
\bibitem{Accardi:2016qay}
Accardi A, et~al.
\newblock \textit{Phys. Rev.} D93:114017 (2016)\relax
\relax
\bibitem{Sato:2016tuz}
Sato N, et~al.
\newblock \textit{Phys. Rev.} D93:074005 (2016)\relax
\relax
\bibitem{Ball:2009mk}
Ball RD, et~al.
\newblock \textit{Nucl. Phys.} B823:195 (2009)\relax
\relax
\bibitem{Martin:2012da}
Martin AD, et~al.
\newblock \textit{Eur. Phys. J.} C73:2318 (2013)\relax
\relax
\bibitem{Ball:2018odr}
Ball RD, Deshpande A.
\newblock In \textit{From My Vast Repertoire ...: Guido Altarelli's Legacy},
  eds. A~Levy, S~Forte, G~Ridolfi.  205--226 arXiv:1801.04842 [hep-ph]
  (2019)\relax
\relax
\bibitem{Ball:2018twp}
Ball RD, Nocera ER, Pearson RL.
\newblock \textit{Eur. Phys. J.} C79:282 (2019)\relax
\relax
\bibitem{AbdulKhalek:2019bux}
Abdul~Khalek R, et~al.
\newblock \textit{Eur. Phys. J.} C:79:838 (2019)\relax
\relax
\bibitem{AbdulKhalek:2019ihb}
Abdul~Khalek R, et~al.
\newblock \textit{Eur. Phys. J.} C79:931 (2019)\relax
\relax
\bibitem{Alekhin:2014irh}
Alekhin S, et~al.
\newblock \textit{Eur. Phys. J.} C75:304 (2015)\relax
\relax
\bibitem{Buckley:2014ana}
Buckley A, et~al.
\newblock \textit{Eur. Phys. J.} C75:132 (2015)\relax
\relax
\bibitem{Carrazza:2014gfa}
Carrazza S, Ferrara A, Palazzo D, Rojo J.
\newblock \textit{J. Phys.} G42:057001 (2015)\relax
\relax
\bibitem{Hou:2019efy}
Hou TJ, et~al.  arXiv:1912.10053 [hep-ph] (2019)\relax
\relax
\bibitem{Sato:2019yez}
Sato N, Andres C, Ethier JJ, Melnitchouk W  arXiv:1905.03788 [hep-ph]
  (2019)\relax
\relax
\bibitem{Leader:2014uua}
Leader E, Sidorov AV, Stamenov DB.
\newblock \textit{Phys. Rev.} D91:054017 (2015)\relax
\relax
\bibitem{deFlorian:2019zkl}
De~Florian D, et~al.
\newblock \textit{Phys. Rev.} D100:114027 (2019)\relax
\relax
\bibitem{Nocera:2014gqa}
Nocera ER, et~al.
\newblock \textit{Nucl. Phys.} B887:276 (2014)\relax
\relax
\bibitem{Walt:2019slu}
Walt M, Helenius I, Vogelsang W.
\newblock \textit{Phys. Rev.} D100:096015 (2019)\relax
\relax
\bibitem{Alekhin:2018pai}
Alekhin S, Blümlein J, Moch S.
\newblock \textit{Eur. Phys. J.} C78:477 (2018)\relax
\relax
\bibitem{Harland-Lang:2016yfn}
Harland-Lang LA, Martin AD, Motylinski P, Thorne RS.
\newblock \textit{Eur. Phys. J.} C76:186 (2016)\relax
\relax
\bibitem{Bailey:2019yze}
Bailey S, Harland-Lang L  arXiv:1909.10541 [hep-ph] (2019)\relax
\relax
\bibitem{Harland-Lang:2017ytb}
Harland-Lang LA, Martin AD, Thorne RS.
\newblock \textit{Eur. Phys. J.} C78:248 (2018)\relax
\relax
\bibitem{Harland-Lang:2015nxa}
Harland-Lang LA, Martin AD, Motylinski P, Thorne RS.
\newblock \textit{Eur. Phys. J.} C75:435 (2015)\relax
\relax
\bibitem{Harland-Lang:2019pla}
Harland-Lang LA, Martin AD, Nathvani R, Thorne RS.
\newblock \textit{Eur. Phys. J.} C79:811 (2019)\relax
\relax
\bibitem{Campbell:2018wfu}
Campbell JM, Rojo J, Slade E, Williams C.
\newblock \textit{Eur. Phys. J.} C78:470 (2018)\relax
\relax
\bibitem{Nocera:2019wyk}
Nocera ER, Ubiali M, Voisey C  arXiv:1912.09543 [hep-ph] (2019)\relax
\relax
\bibitem{Ball:2018iqk}
Ball RD, et~al.
\newblock \textit{Eur. Phys. J.} C78:408 (2018)\relax
\relax
\bibitem{Ball:2017otu}
Ball RD, et~al.
\newblock \textit{Eur. Phys. J.} C78:321 (2018)\relax
\relax
\bibitem{Jimenez-Delgado:2014twa}
Jimenez-Delgado P, Reya E.
\newblock \textit{Phys. Rev.} D89:074049 (2014)\relax
\relax
\bibitem{Abramowicz:2015mha}
Abramowicz H, et~al.
\newblock \textit{Eur. Phys. J.} C75:580 (2015)\relax
\relax
\bibitem{Diehl:2019fsz}
Diehl M, Stienemeier P  arXiv:1904.10722 [hep-ph] (2019)\relax
\relax
\bibitem{Kassabov:2019xxx}
 See Z.~Kassabov talk presented at the workshop {\it Ultimate Precision at
  Hadron Colliders}, 25 Nov. - 6 Dec. 2019, Institut Pascal, Orsay\relax
\relax
\bibitem{Borsa:2017vwy}
Borsa I, Sassot R, Stratmann M.
\newblock \textit{Phys. Rev.} D96:094020 (2017)\relax
\relax
\bibitem{Bonvini:2015ira}
Bonvini M, et~al.
\newblock \textit{JHEP} 09:191 (2015)\relax
\relax
\bibitem{Beenakker:2015rna}
Beenakker W, et~al.
\newblock \textit{Eur. Phys. J.} C76:53 (2016)\relax
\relax
\bibitem{Martin:2004dh}
Martin AD, Roberts RG, Stirling WJ, Thorne RS.
\newblock \textit{Eur. Phys. J.} C39:155 (2005)\relax
\relax
\bibitem{Schmidt:2015zda}
Schmidt C, Pumplin J, Stump D, Yuan CP.
\newblock \textit{Phys. Rev.} D93:114015 (2016)\relax
\relax
\bibitem{Ball:2013hta}
Ball RD, et~al.
\newblock \textit{Nucl. Phys.} B877:290 (2013)\relax
\relax
\bibitem{deFlorian:2014yva}
de~Florian D, Sassot R, Stratmann M, Vogelsang W.
\newblock \textit{Phys. Rev. Lett.} 113:012001 (2014)\relax
\relax
\bibitem{Adamczyk:2016okk}
Adamczyk L, et~al.
\newblock \textit{Phys. Rev.} D95:071103 (2017)\relax
\relax
\bibitem{Adam:2018pns}
Adam J, et~al.
\newblock \textit{Phys. Rev.} D98:032011 (2018)\relax
\relax
\bibitem{Nocera:2017wep}
Nocera ER. 2017.
\newblock In \textit{{22nd International Symposium on Spin Physics (SPIN 2016)
  Urbana, IL, USA, September 25-30, 2016}}\relax
\relax
\bibitem{Airapetian:2004zf}
Airapetian A, et~al.
\newblock \textit{Phys. Rev.} D71:012003 (2005)\relax
\relax
\bibitem{Alekseev:2009ac}
Alekseev M, et~al.
\newblock \textit{Phys. Lett.} B680:217 (2009)\relax
\relax
\bibitem{Alekseev:2010ub}
Alekseev MG, et~al.
\newblock \textit{Phys. Lett.} B693:227 (2010)\relax
\relax
\bibitem{Adamczyk:2012qj}
Adamczyk L, et~al.
\newblock \textit{Phys. Rev.} D86:032006 (2012)\relax
\relax
\bibitem{Adamczyk:2014ozi}
Adamczyk L, et~al.
\newblock \textit{Phys. Rev. Lett.} 115:092002 (2015)\relax
\relax
\bibitem{Adare:2008qb}
Adare A, et~al.
\newblock \textit{Phys. Rev.} D79:012003 (2009)\relax
\relax
\bibitem{Adare:2008aa}
Adare A, et~al.
\newblock \textit{Phys. Rev. Lett.} 103:012003 (2009)\relax
\relax
\bibitem{Adare:2014hsq}
Adare A, et~al.
\newblock \textit{Phys. Rev.} D90:012007 (2014)\relax
\relax
\bibitem{Adolph:2012ca}
Adolph C, et~al.
\newblock \textit{Phys. Rev.} D87:052018 (2013)\relax
\relax
\bibitem{Adamczyk:2014xyw}
Adamczyk L, et~al.
\newblock \textit{Phys. Rev. Lett.} 113:072301 (2014)\relax
\relax
\bibitem{Stratmann:2020xxx}
 See Stratmann M and Surrow B, Recent Progress on the Gluon Polarization, this
  volume\relax
\relax
\bibitem{Kusina:2016fxy}
Kusina A, et~al.
\newblock \textit{Eur. Phys. J.} C77:488 (2017)\relax
\relax
\bibitem{Eskola:2019dui}
Eskola KJ, Paakkinen P, Paukkunen H.
\newblock \textit{Eur. Phys. J.} C79:511 (2019)\relax
\relax
\bibitem{Eskola:2019bgf}
Eskola KJ, Helenius I, Paakkinen P, Paukkunen H  arXiv:1906.02512 [hep-ph]
  (2019)\relax
\relax
\bibitem{Aubert:1983xm}
Aubert JJ, et~al.
\newblock \textit{Phys. Lett.} 123B:275 (1983)\relax
\relax
\bibitem{Aad:2012tfa}
Aad G, et~al.
\newblock \textit{Phys. Lett.} B716:1 (2012)\relax
\relax
\bibitem{Chatrchyan:2012xdj}
Chatrchyan S, et~al.
\newblock \textit{Phys. Lett.} B716:30 (2012)\relax
\relax
\bibitem{Tanabashi:2018oca}
Tanabashi M, et~al.
\newblock \textit{Phys. Rev.} D98:030001 (2018)\relax
\relax
\bibitem{Forte:2020pyp}
Forte S, Kassabov Z  arXiv:2001.04986 [hep-ph] (2020)\relax
\relax
\bibitem{Cepeda:2019klc}
Cepeda M, et~al.
\newblock \textit{CERN Yellow Rep. Monogr.} 7:221 (2019)\relax
\relax
\bibitem{Azzi:2019yne}
Azzi P, et~al.
\newblock \textit{CERN Yellow Rep. Monogr.} 7:1 (2019)\relax
\relax
\bibitem{Bagnaschi:2019mzi}
Bagnaschi E, Vicini A  arXiv:1910.04726 [hep-ph] (2019)\relax
\relax
\bibitem{Brivio:2017vri}
Brivio I, Trott M.
\newblock \textit{Phys. Rept.} 793:1 (2019)\relax
\relax
\bibitem{Hartland:2019bjb}
Hartland NP, et~al.
\newblock \textit{JHEP} 04:100 (2019)\relax
\relax
\bibitem{Carrazza:2019sec}
Carrazza S, et~al.
\newblock \textit{Phys. Rev. Lett.} 123:132001 (2019)\relax
\relax
\bibitem{Amaudruz:1991at}
Amaudruz P, et~al.
\newblock \textit{Phys. Rev. Lett.} 66:2712 (1991)\relax
\relax
\bibitem{Towell:2001nh}
Towell RS, et~al.
\newblock \textit{Phys. Rev.} D64:052002 (2001)\relax
\relax
\bibitem{Melnitchouk:1991ui}
Melnitchouk W, Thomas AW, Signal AI.
\newblock \textit{Z. Phys.} A340:85 (1991)\relax
\relax
\bibitem{Kumano:1991em}
Kumano S, Londergan JT.
\newblock \textit{Phys. Rev.} D44:717 (1991)\relax
\relax
\bibitem{Henley:1990kw}
Henley EM, Miller GA.
\newblock \textit{Phys. Lett.} B251:453 (1990)\relax
\relax
\bibitem{Speth:1996pz}
Speth J, Thomas AW.
\newblock \textit{Adv. Nucl. Phys.} 24:83 (1997), [,83(1996)]\relax
\relax
\bibitem{Ethier:2018efr}
Ethier JJ, Melnitchouk W, Steffens F, Thomas AW.
\newblock \textit{Phys. Rev.} D100:034014 (2019)\relax
\relax
\bibitem{Aidala:2017ofy}
Aidala CA, et~al.
\newblock \textit{Nucl. Instrum. Meth.} A930:49 (2019)\relax
\relax
\bibitem{Nocera:2014rea}
Nocera ER.
\newblock \textit{PoS} DIS2014:204 (2014)\relax
\relax
\bibitem{Accardi:2011fa}
Accardi A, et~al.
\newblock \textit{Phys. Rev.} D84:014008 (2011)\relax
\relax
\bibitem{Owens:2012bv}
Owens JF, Accardi A, Melnitchouk W.
\newblock \textit{Phys. Rev.} D87:094012 (2013)\relax
\relax
\bibitem{Arrington:2011qt}
Arrington J, Rubin JG, Melnitchouk W.
\newblock \textit{Phys. Rev. Lett.} 108:252001 (2012)\relax
\relax
\bibitem{Tropiano:2018quk}
Tropiano AJ, Ethier JJ, Melnitchouk W, Sato N.
\newblock \textit{Phys. Rev.} C99:035201 (2019)\relax
\relax
\bibitem{Dudek:2012vr}
Dudek J, et~al.
\newblock \textit{Eur. Phys. J.} A48:187 (2012)\relax
\relax
\bibitem{MARATHON}
Petratos GG, et~al. Jefferson Lab PAC37 Proposal (2010), experiment
  E12-10-103\relax
\relax
\bibitem{Brodsky:2015fna}
Brodsky SJ, et~al.
\newblock \textit{Adv. High Energy Phys.} 2015:231547 (2015)\relax
\relax
\bibitem{Dulat:2013hea}
Dulat S, et~al.
\newblock \textit{Phys. Rev.} D89:073004 (2014)\relax
\relax
\bibitem{Jimenez-Delgado:2014zga}
Jimenez-Delgado P, Hobbs TJ, Londergan JT, Melnitchouk W.
\newblock \textit{Phys. Rev. Lett.} 114:082002 (2015)\relax
\relax
\bibitem{Leader:2013jra}
Leader E, Lorc\'e C.
\newblock \textit{Phys. Rept.} 541:163 (2014)\relax
\relax
\bibitem{Hen:2013oha}
Hen O, et~al.
\newblock \textit{Int. J. Mod. Phys.} E22:1330017 (2013)\relax
\relax
\bibitem{Abreu:2007kv}
Abreu S, et~al.
\newblock \textit{J. Phys.} G35:054001 (2008)\relax
\relax
\bibitem{Adams:2005dq}
Adams J, et~al.
\newblock \textit{Nucl. Phys.} A757:102 (2005)\relax
\relax
\bibitem{Acharya:2019vlb}
Acharya S, et~al.
\newblock \textit{Phys. Lett.} B798:134926 (2019)\relax
\relax
\bibitem{Adam:2019rxb}
Adam J.
\newblock \textit{PoS} DIS2019:042 (2019)\relax
\relax
\bibitem{Ryskin:1992ui}
Ryskin MG.
\newblock \textit{Z. Phys.} C57:89 (1993)\relax
\relax
\bibitem{IceCube:2018cha}
Aartsen MG, et~al.
\newblock \textit{Science} 361:147 (2018)\relax
\relax
\bibitem{Adrian-Martinez:2016fdl}
Adrian-Martinez S, et~al.
\newblock \textit{J. Phys.} G43:084001 (2016)\relax
\relax
\bibitem{Avrorin:2018ijk}
Avrorin AD, et~al.
\newblock \textit{EPJ Web Conf.} 191:01006 (2018)\relax
\relax
\bibitem{Aab:2019gra}
Aab A, et~al.
\newblock \textit{Front. Astron. Space Sci.} 6:24 (2019)\relax
\relax
\bibitem{Bertone:2018dse}
Bertone V, Gauld R, Rojo J.
\newblock \textit{JHEP} 01:217 (2019)\relax
\relax
\bibitem{Bhattacharya:2016jce}
Bhattacharya A, et~al.
\newblock \textit{JHEP} 11:167 (2016)\relax
\relax
\bibitem{Klein:2020nuk}
Klein SR, Robertson SA, Vogt R  arXiv:2001.03677 [hep-ph] (2020)\relax
\relax
\bibitem{Henley:2005ms}
Henley EM, Jalilian-Marian J.
\newblock \textit{Phys. Rev.} D73:094004 (2006)\relax
\relax
\bibitem{Apollinari:2017cqg}
Apollinari G, Brüning O, Nakamoto T, Rossi L.
\newblock \textit{CERN Yellow Rep.} :1 (2015)\relax
\relax
\bibitem{AbelleiraFernandez:2012cc}
Abelleira~Fernandez JL, et~al.
\newblock \textit{J. Phys.} G39:075001 (2012)\relax
\relax
\bibitem{Accardi:2012qut}
Accardi A, et~al.
\newblock \textit{Eur. Phys. J.} A52:268 (2016)\relax
\relax
\bibitem{Boer:2011fh}
Boer D, et~al.  arXiv:1108.1713 [nucl-th] (2011)\relax
\relax
\bibitem{AbdulKhalek:2019mps}
Abdul~Khalek R, et~al.
\newblock \textit{SciPost Phys.} 7:051 (2019)\relax
\relax
\bibitem{Khalek:2018mdn}
Abdul~Khalek R, et~al.
\newblock \textit{Eur. Phys. J.} C78:962 (2018)\relax
\relax
\bibitem{Aschenauer:2019kzf}
Aschenauer EC, Borsa I, Sassot R, Van~Hulse C.
\newblock \textit{Phys. Rev.} D99:094004 (2019)\relax
\relax
\bibitem{Aschenauer:2012ve}
Aschenauer EC, Sassot R, Stratmann M.
\newblock \textit{Phys. Rev.} D86:054020 (2012)\relax
\relax
\bibitem{Aschenauer:2015ata}
Aschenauer EC, Sassot R, Stratmann M.
\newblock \textit{Phys. Rev.} D92:094030 (2015)\relax
\relax
\bibitem{Aschenauer:2013iia}
Aschenauer EC, et~al.
\newblock \textit{Phys. Rev.} D88:114025 (2013)\relax
\relax
\bibitem{Ball:2013tyh}
Ball RD, et~al.
\newblock \textit{Phys. Lett.} B728:524 (2014)\relax
\relax
\bibitem{HannuPaukkunenfortheLHeCstudygroup:2017ric}
Paukkunen H.
\newblock \textit{PoS} DIS2017:109 (2018)\relax
\relax
\bibitem{Helenius:2015xda}
Helenius I, Paukkunen H, Armesto N.
\newblock \textit{PoS} DIS2015:226 (2015)\relax
\relax
\bibitem{Aschenauer:2017oxs}
Aschenauer EC, et~al.
\newblock \textit{Phys. Rev.} D96:114005 (2017)\relax
\relax
\bibitem{Lin:2017snn}
Lin HW, et~al.
\newblock \textit{Prog. Part. Nucl. Phys.} 100:107 (2018)\relax
\relax
\bibitem{Angeles-Martinez:2015sea}
Angeles-Martinez R, et~al.
\newblock \textit{Acta Phys. Polon.} B46:2501 (2015)\relax
\relax
\bibitem{Bacchetta:2016ccz}
Bacchetta A.
\newblock \textit{Eur. Phys. J.} A52:163 (2016)\relax
\relax
\end{mcbibliography}

\end{document}